\documentclass[11pt,a4paper]{article} 
 
\usepackage[utf8]{inputenc}
\usepackage{fontenc}
\usepackage[english]{babel}
\usepackage{amsmath}
\usepackage{geometry}
\usepackage{amsthm}
\usepackage{mathrsfs}
\usepackage{bbold}
\usepackage{mathtools}
\usepackage{amsfonts}
\usepackage{graphicx}
\usepackage{physics}
\usepackage{comment}
\usepackage{bbm}
\usepackage{layout}
\usepackage{xcolor}
\usepackage{dsfont}	
\usepackage{slashed}
\usepackage{amssymb}
	
\usepackage{jheppub} 

\usepackage{natbib}
\author[a]{Luca Griguolo,}
\author[b]{Luigi Guerrini,}
\author[c]{Rodolfo Panerai,}
\author[d]{Jacopo Papalini,}
\author[b]{and Domenico Seminara}
\affiliation[a]{Dipartimento SMFI, Universit\`a di Parma and INFN Gruppo Collegato di Parma, \\ Viale G.P. Usberti 7/A, 43100 Parma, Italy}
\affiliation[b]{Dipartimento di Fisica, Universit\`a di Firenze and INFN Sezione di Firenze, \\ via G. Sansone 1, 50019 Sesto Fiorentino, Italy} 
\affiliation[c]{Institute for Theoretical Physics, University of Cologne, \\ Z\"ulpicher Straße 77, 50937 K\"oln, Germany}
\affiliation[d]{Galileo Galilei Institute for Theoretical Physics, INFN, \\ Largo Enrico Fermi, 2, 50125 Firenze, Italy}

\title{Supersymmetric localization of (higher-spin) \\ JT gravity: a bulk perspective}

\abstract{We study two-dimensional Jackiw--Teitelboim gravity on the disk topology by using a BF gauge theory in the presence of a boundary term. The system can be equivalently written in a supersymmetric way by introducing auxiliary gauginos and scalars with suitable boundary conditions on the hemisphere. We compute the exact partition function thanks to supersymmetric localization and we recover the result obtained from the Schwarzian theory by accurately identifying the physical scales. The calculation is then easily extended to the higher-spin generalization of Jackiw--Teitelboim gravity, finding perfect agreement with previous results. We argue that our procedure can also be applied to boundary-anchored Wilson lines correlators.}

\newcommand{\Q}{\mathbb{Q}}
\newcommand{\e}{\mathrm{e}}

\begin{document}
\maketitle

\section{Introduction}
The AdS/CFT correspondence \cite{Maldacena:1997re,Witten:1998qj,Gubser:1998bc} represents a promising framework to understand and, hopefully, to solve some subtle problems related to the quantization of gravity. Through the correspondence, the boundary theory can serve as a guide for understanding properties of the bulk physics. This is especially useful given the notorious difficulties in making sense of the functional integral of quantum gravity.

A powerful non-perturbative method to perform exact computations in certain quantum field theories is the localization technique \cite{Pestun:2016zxk}, where the functional integral can be shown to ``localize'' over some solutions in field space, parameterizing a moduli space of suitable classical configurations. In simple cases, this finite-dimensional integral can be evaluated analytically, leading to a complete solution of the problem. When a system (or a particular set of observables) having a dual gravitational description in a bulk space, can be studied exactly through localization, we would expect to learn something about the structure of the related quantum gravity path integral. More ambitiously, we also hope that the bulk theory inherits some localization properties, opening to the possibility of obtaining exact results for integrations on fluctuating backgrounds.

The program of studying gravitational systems from localization techniques applied to the boundary theory has been successfully exploited to derive the Bekenstein-Hawking entropy of supersymmetric black holes in AdS$_4$ \cite{Benini:2015noa,Benini:2015eyy,Azzurli:2017kxo} and AdS$_5$ \cite{Cabo-Bizet:2018ehj,Choi:2018hmj,Benini:2018ywd}. There have also been attempts to extend these localization methods directly to supergravity to evaluate the bulk quantum gravity partition function, mainly in the context of AdS$_4$/CFT$_{3}$ holography \cite{Dabholkar:2014wpa,Murthy:2015yfa}.

On general grounds, one expects that, in low dimensions, our understanding of quantum gravity should improve. In particular, 2d/1d holography provides probably the best frameworks to study fluctuating geometries beyond perturbation theory. Gravitons or gauge bosons in two dimensions have no dynamical degrees of freedom; therefore, the quantum path integral in general simplifies dramatically, even in non-supersymmetric settings. On the other hand, many of the open questions from higher dimensional holography, such as bulk reconstruction or the physics of black holes and wormholes, persist in the lowest dimensional case.

Two-dimensional Jackiw--Teitelboim (JT) theory \cite{Teitelboim:1983ux,Jackiw:1984je} involving, in the second-order formalism, a dilaton field $\Phi$ and the metric tensor $\mathrm{g}_{\mu\nu}$, is a simple tractable example of holographic correspondence and has attracted much attention in the last few years (see \cite{Mertens:2022irh} for a recent review). The dual holographic theory is a one-dimensional theory, the Schwarzian quantum mechanics \cite{Maldacena:2016upp}, that effectively describes the bulk quantum gravity: a large class of correlation functions is precisely mapped to the boundary theory. Quite interestingly, the Schwarzian path integral can be exactly evaluated thanks to equivariant localization \cite{Stanford:2017thb}, leading to a precise expression for the thermal-JT partition function at the basis of many recent developments \cite{Kitaev:2017awl,Saad:2018bqo,Stanford:2019vob,Penington:2019kki,Almheiri:2019qdq}. In particular, higher-genus contributions to the path integral play a fundamental role in deriving a non-perturbative extension of the holographic duality in terms of an ensemble of theories described by a double-scaled matrix model \cite{Saad:2019lba}.

This paper proposes that the same results can be obtained starting from the bulk theory and using a supersymmetric localization procedure \cite{Benini:2012ui}. The main idea\footnote{In three-dimensional gravity there have been similar attempts, both for the AdS \cite{Iizuka:2015jma} and the dS \cite{Castro:2023bvo} cases} behind our computation is to use the well-known formulation of JT gravity as a BF gauge theory based on the algebra of $\mathrm{SL}(2,\mathbb{R})$ \cite{Blommaert:2018oro,Mertens:2018fds,Iliesiu:2019xuh}. We map the theory into a supersymmetric $\mathcal{N}=\left(2,2\right)$ gauge theory on the hemisphere and apply supersymmetric localization to reproduce the known Schwarzian partition function.

A subtle point concerns the correct identification of the physical scales present in the gravitational theory with the geometrical scales appearing in the supersymmetric BF theory. We argue that our gauge model is actually obtained by reducing the three-dimensional Chern--Simons theory on a solid torus conformally equivalent to thermal AdS$_3$ and this provides us with the correct supersymmetric boundary terms and scales identification.

Interestingly our approach is easily extended for the higher-spin generalization of JT gravity \cite{Alkalaev:2014qpa,Gonzalez:2018enk}. In this case, the relevant gauge theory is a $\mathrm{SL}(N,\mathbb{R})$ BF theory, and its partition function has been derived using equivariant localization on the boundary $\mathrm{SL}(N,\mathbb{R})$ Schwarzian quantum mechanics \cite{Gonzalez:2018enk, Datta:2021efl, Kruthoff:2022voq}. 
Our procedure reproduces precisely the result of \cite{Gonzalez:2018enk,Kruthoff:2022voq}, bypassing the technical complications related to the derivation of the boundary quantum theory.

Obtaining the JT partition function through localization is fascinating as it suggests that we can use the same framework to compute general observables. In the gauge theory formulation, correlation functions of boundary-anchored Wilson lines \cite{Blommaert:2018oro} are the most natural candidates to be studied. Physically, they represent correlators of bi-local operators in the Schwarzian theory and contain essential information about the quantum structure of the bulk gravity. While bi-local correlators have been thoroughly studied in standard JT gravity \cite{Mertens:2017mtv,Blommaert:2018oro,Iliesiu:2019xuh} obtaining exact expressions from different methods, their higher-spin cousins have never been considered. Supersymmetric localization could provide a convenient framework for their calculation.

The structure of the paper is the following. We start Section~\ref{SEC:JT_to_BF} by reviewing the gauge formulation of JT gravity, and then proceed to construct an equivalent supersymmetric BF theory. We pay particular attention to imposing supersymmetric boundary conditions on the hemisphere and to elucidate the identification between the physical and the geometrical scales. The actual localization of the path-integral is performed in Section~\ref{localization}: we present the localizing term and compute the relevant functional determinants obtaining the well-known final result for the JT partition function. In Section~\ref{higher}, we extend the computation to the higher-spin generalization of JT gravity recovering the disk partition function obtained in \cite{Gonzalez:2018enk,Kruthoff:2022voq}. Section~\ref{concl} contains our conclusions and speculations about further uses of our procedure. The paper is completed with a couple of technical appendices.

\section{JT gravity as a supersymmetric BF theory}\label{SEC:JT_to_BF}
\subsection{JT gravity as a BF theory}\label{sec1}
Let us start by briefly reviewing how JT gravity can be formulated as a two-dimensional BF theory with $\mathrm{SL}(2,\mathbb{R})$ gauge group \cite{Fukuyama:1985gg,Isler:1989hq,Chamseddine:1989yz,Iliesiu:2019xuh}. In particular, our focus is on the theory defined on a two-dimensional manifold $\Sigma$ with the topology of the disk. We start from the BF action
\begin{equation}\label{act}
	S_{\mathrm{BF}} = -i \int_{\Sigma} \Tr(\chi F) \;,
\end{equation}
where $F = \mathrm{d}A-A \wedge A$ is the field strength associated with a gauge connection one form $A$, and $\chi$ is an auxiliary scalar field in the adjoint representation of the gauge group.

We consider a basis $\{\mathsf{P}_0,\mathsf{P}_1,\mathsf{P}_2\}$ for the generators of $\mathfrak{sl}(2,\mathbb{R})$ obeying the commutation relations
\begin{align}\label{cmt2}
	[\mathsf{P}_0,\mathsf{P}_1] &= \mathsf{P}_{2} \;, & [\mathsf{P}_0,\mathsf{P}_2] &= -\mathsf{P}_{1} \;, & [\mathsf{P}_1,\mathsf{P}_2] &= -\mathsf{P}_{0} \;.
\end{align}
This algebra can be explicitly realized by choosing, for instance, a real two-dimensional representation in terms of Pauli matrices with
\begin{align}\label{cmt}
	\mathsf{P}_0 &= \frac{i\sigma_2}{2} \;, & \mathsf{P}_1 &= \frac{\sigma_1}{2} \;, & \mathsf{P}_2 &= \frac{\sigma_3}{2} \;.
\end{align}
The corresponding Killing form reads $\Tr(\mathsf{P}_i \mathsf{P}_j) = \frac{1}{2}\,\mathrm{diag}(-1,+1,+1)$.

We then expand the fields on such a basis as
\begin{align}\label{fields1}
	A &= \sqrt{\frac{\Lambda}{2}} \, e^{a} \, \mathsf{P}_{a} + \omega \, \mathsf{P}_0 \;, \\
	\chi &= \chi^{a} \, \mathsf{P}_{a} + \chi^0 \, \mathsf{P}_0 \;.
\end{align}
Here, we regard the index $a\in{1,2}$ as an $\mathrm{SO}(2)$ frame index.
In fact, the matching of gravitational and gauge degrees of freedom is obtained by identifying the one-forms $e^{a}$ and $\omega$ with the zweibein and the spin connection, respectively. In the spirit of the first-order formulation of gravity, these are regarded as independent degrees of freedom.

Exploiting the expansion \eqref{fields1} and the commutation relations \eqref{cmt2}, we compute the nonabelian field strength
\begin{align}
	F
	&= F^{a}\,\mathsf{P}_a + F^{0}\,\mathsf{P}_{0} \cr
	&= \sqrt{\frac{\Lambda}{2}} \, \bigg(\mathrm{d}e^{a} + \epsilon^{a}{}_{b} \, \omega \wedge e^b\bigg) \, \mathsf{P}_a + \bigg(\mathrm{d}\omega+\frac{\Lambda}{4} \, \epsilon_{ab} \, e^{a}\wedge e^b\bigg) \, \mathsf{P}_0 \;.
\end{align}
Upon plugging this expression into the BF action \eqref{act} one finds that the variation of $S_{\mathrm{BF}}$ with respect to the $\mathrm{SO}(2)$ vector $\chi^a$ yields the equations of motion
\begin{equation}\label{torsion}
	\mathrm{d}e^{a} + \epsilon^{a}{}_{b} \, \omega \wedge e^b = 0 \;.
\end{equation}
This is precisely the zero-torsion condition, which, once solved, gives the spin connection $\omega$ in terms of the zweibein $e^a$. The action \eqref{act} evaluated on the solutions of \eqref{torsion} reduces to the second-order action
\begin{equation}\label{bf2}
	S_{\mathrm{BF}} = \frac{i}{2} \int_{\Sigma} \chi_0 \left(\mathrm{d} \omega(e) +\frac{\Lambda}{2} \, e^{1} \wedge e^2\right) \;.
\end{equation}
In two dimensions, $\mathrm{d}\omega = R/2 \; e^1 \wedge e^2$, and we recognize in $e^1 \wedge e^2$ the two-dimensional volume form. We can then rewrite \eqref{bf2} in terms of the metric $g = \delta_{ab} \, e^a \otimes e^b$ as
\begin{equation}\label{bf4}
	S_{\mathrm{BF}} = \frac{i}{4} \int_{\Sigma} \mathrm{d}^2x \, \sqrt{g} \, \chi_0  \, (R(g) +\Lambda) \;.
\end{equation}
In \eqref{bf4}, $S_{\mathrm{BF}}$ reproduces the bulk contribution of the JT action if we identify the dilaton field with $\Phi = -i \, \chi_0 /4$.

At this point, within the gauge formulation of JT gravity, it is common practice to introduce a boundary term that, when combined with appropriate boundary conditions, replicates the dynamics of the Schwarzian theory on the boundary of the disk \cite{Blommaert:2018oro,Mertens:2018fds,Iliesiu:2019xuh}. In the metric formulation, this consists in implementing the Gibbons--Hawking boundary term. However, we will take a different approach here, utilizing supersymmetry. The process of supersymmetrization will naturally guide us towards incorporating a suitable boundary term that, upon identifying the correct physical scales, will allow us to obtain the Schwarzian partition function. In the subsequent sections, we will provide a more detailed explanation of these steps.

\subsection{Supersymmetrizing JT gravity}\label{sec2}
As a next step, we will introduce new auxiliary degrees of freedom in the BF action \eqref{act} with the aim of making it supersymmetric.
In doing so, we first introduce a Riemannian structure on $\Sigma$.\footnote{
	The reader should not confuse the dynamical geometry associated with the degrees of freedom of the gauge theory with the background geometry introduced to construct the supersymmetry algebra.
} We identify $\Sigma$ with the hemisphere $\mathrm{HS}^2$ endowed with the metric 
\begin{equation}\label{EQ:hemisphere_metric}
	\mathrm{d}s^2 = \ell^2\left(\mathrm{d}\theta^2 + \sin^2\!\theta \; \mathrm{d}\varphi^2\right) \;,
\end{equation}
written in terms of conventional spherical coordinates $\theta\in\left[0,\pi/2\right]$ and $\varphi\in[0,2\pi)$. The boundary circle is located at $\theta=\pi/2$.

A simple scheme for localizing gauge theories on $\mathrm{HS}^2$ was developed in \cite{Sugishita:2013jca, Honda:2013uca, Hori:2013ika} in the presence of $\mathcal{N}=(2,2)$ supersymmetry. In order to leverage such results, we need to embed the degrees of freedom of the BF theory into an $\mathcal{N}=(2,2)$ vector multiplet. This off-shell multiplet contains a two-dimensional gauge connection $A$, two scalars $\eta$ and $\sigma$ of dimension one, two Dirac fermions $\lambda$ and $\bar\lambda$, and an auxiliary field $D$. The associated supersymmetry variations, parametrized by conformal Killing spinors $\epsilon$ and $\bar\epsilon$, are given in Appendix~\ref{SUSYonHS2}, where the geometry and the supersymmetry of the hemisphere are spelled out in detail.

While it is straightforward to identify the gauge field in the vector multiplet with the one appearing in the BF action, we have two options for the scalar field $\chi$, namely $\sigma$ and $\eta$. In choosing between them, we recall that the BF action can be constructed by dimensionally-reducing the Chern--Simons action. In this framework, the scalar $\chi$ can be identified with the third component of the gauge field in three dimensions. Similarly, the $\mathcal{N}=(2,2)$ vector multiplet can be obtained by performing a dimensional reduction of the $\mathcal{N}=2$ vector multiplet in three dimensions, $\sigma$ originates from the third component of the vector field. To address the disparity in dimensions between the dimensionless field $\chi$ and the dimensionful $\sigma$ we set
\begin{equation}\label{rescaling}
	\chi = L \sigma
\end{equation}
where $L$ is a generic length scale. We will temporarily withhold any assumption about $L$, which will be determined in Section~\ref{scales}.

With the identification \eqref{rescaling}, the BF action \eqref{act} reads
\begin{align}\label{EQ:S_BF_with_sigma}
	S_{\textup{BF}} = -iL\int_{\mathrm{HS}^2} \mathrm{d}^2x \, \sqrt{g} \, \Tr(\sigma f) \;,
\end{align}
where the scalar $f = \star F$ is the Hodge dual of the field strength two-form.

The above action has its supersymmetric completion in the bulk action
\begin{align}\label{extended_BF}
	S_{\textup{bulk}} = -iL\int_{\mathrm{HS}^2} \mathrm{d}^2x \, \sqrt{g} \, \Tr(\sigma f-\frac{1}{2}\bar\lambda\lambda+ D \eta) \;,
\end{align}
which is still equivalent to \eqref{EQ:S_BF_with_sigma} since the additional degrees of freedom are non-dynamical.

For the hemisphere, however, the supersymmetric variation of \eqref{extended_BF} produces a boundary term originating from the integration of a total divergence, namely
\begin{align}\label{EQ:delta_S_bulk}
	\delta_{\epsilon,\bar\epsilon} S_{\textup{bulk}}
	&= \frac{L}{2}\int_{\mathrm{HS}^2} \mathrm{d}^2x \, \sqrt{g} \,\Tr(D_{\mu}[\eta(\bar\lambda\gamma^\mu\epsilon-\bar\epsilon\gamma^\mu\lambda)-\sigma\epsilon^{\mu\nu}(\bar\epsilon\gamma_\nu\lambda+\bar\lambda\gamma_\nu\epsilon)]) \cr
	&= -iL \oint_{\partial\mathrm{HS}^2} \mathrm{d}\varphi \; \Tr(\frac{i}{2}\ell^2\eta(\bar\lambda\gamma^\theta\epsilon-\bar\epsilon\gamma^\theta\lambda)+\sigma\delta_{\epsilon,\bar{\epsilon}} A_\varphi) \;.
\end{align}
In order to obtain a supersymmetric action we should then complement $S_{\mathrm{bulk}}$ with the boundary term
\begin{equation}\label{boundary}
	S_{\textup{bdry}} = L \ell \oint_{\partial\mathrm{HS}^2} \mathrm{d}\varphi \; \Tr(\sigma^2) \;.
\end{equation}
In fact, for half of the supersymmetry\footnote{Specifically, the preserved supercharges are those generated by \eqref{preservedsusy}.} on $\partial\mathrm{HS}^2$, the second term in \eqref{EQ:delta_S_bulk} is exactly canceled by the supersymmetric variation of \eqref{boundary}, since they nicely combine into  
\begin{equation}
	\delta_{\epsilon,\bar\epsilon} (S_{\textup{bulk}}+S_{\textup{bdry}}) = -iL\oint_{\partial\mathrm{HS}^2} \mathrm{d}\varphi \; \Tr(\sigma\,\delta_{\epsilon,\bar{\epsilon}}(A_\varphi+i\ell\sigma)) + \ldots \;.
\end{equation}
The combination $A_\varphi+i\ell\sigma$ can be regarded as the putative connection of a $1/2$-BPS Wilson loop running along the boundary. The dots stand for a remaining term proportional to $\eta$ coming from \eqref{EQ:delta_S_bulk} that can be eliminated by imposing the boundary condition\footnote{
	The term proportional to $\eta$ in \eqref{EQ:delta_S_bulk} could also be canceled by the variation of an additional boundary contribution proportional to $-i\oint\eta^2$.
} 
\begin{equation}\label{eta_bc}
	\eta\big|_{\partial\mathrm{HS}^2} = 0 \;.
\end{equation}
In summary, we managed to build a supersymmetric version of the BF action, 
\begin{equation}\label{EQ:S_tot}
	S_{\textup{tot}} = L\left[-i\int_{\mathrm{HS}^2} \mathrm{d}^2x\,\sqrt{g}\,\Tr(\sigma f-\frac{1}{2}\bar\lambda\lambda+ D \eta) + \frac{\ell}{2}\oint_{\partial\mathrm{HS}^2} \mathrm{d}\varphi \, \Tr(\sigma^2)\right] \;,
\end{equation}
which preserves half of the off-shell supersymmetry enjoyed by the bulk model on the sphere. We emphasize that, up to this point, both $L$ and $\ell$ are general length scales. We will fix them in the following subsection, by comparison with the relevant scales in JT gravity.

Finally, let us briefly turn our attention to the variational principle associated with the action \eqref{EQ:S_tot}. Upon variation of the fields, we obtain the boundary term
\begin{equation}
	\delta S_{\textup{tot}} = -iL\oint_{\partial {\mathrm{HS}^2}} \mathrm{d}\varphi \; \Tr(\sigma \, \delta( A_{\varphi}+i \ell \sigma)) \;.
\end{equation}
For a well-defined variational principle, this term must vanish. Therefore, we impose the condition:
\begin{equation}\label{var_principle}
	\delta(A_{\varphi} + i \ell \sigma) = 0 \;,
\end{equation}
which implies setting $A_{\varphi} + i\ell\sigma$ equal to a constant. The specific value of this constant will be determined at the end of next section.

\subsection{Gauge and gravitational scales}\label{scales}
At this point, we need to identify the relevant parameters in the BF theory with their gravitational counterparts in order to ensure a precise match between the partition function of JT gravity on the disk topology with that of our supersymmetric theory.

We start by reviewing some well-known facts about three-dimensional gravity with negative cosmological constant. The spectrum of 3d gravity includes global thermal AdS$_3$ and a collection of Euclidean BTZ solutions \cite{Banados:1992wn,Banados:1992gq}, separated from the AdS$_3$ vacuum by a mass gap. All these Euclidean saddles are characterized by the topology of a solid torus. Notably, the modular invariance of 3d gravity naturally acts on the boundary torus of complex structure $\tau$, allowing for their mapping to one another through modular transformations. In particular, a Euclidean geometry with torus boundary is specified once one chooses which cycle of the boundary torus is contractible in the bulk. In the case where the time cycle is contractible, we obtain an Euclidean BTZ solution. On the other hand, when the spatial cycle is contractible, we have thermal AdS$_3$ as the solution. Specifically, the non-rotating BTZ solution is related to thermal AdS$_3$ by a modular $S$-transformation $\tau \rightarrow -1/\tau$, which acts by swapping the two cycles.

Considering 3d coordinates $(t_{\text{E}},r,\phi)$ playing the role of time, radial coordinate, and angular coordinate respectively, it is known that the spherically
symmetric $(t_{\text{E}},r)$ sector of 3d gravity is directly governed by JT gravity \cite{Mertens:2022irh,Mertens:2018fds}. For instance, the JT black hole can be obtained as the dimensional reduction of the BTZ three-dimensional one, by reducing along the circle parametrized by $\phi$. In particular, the inverse temperature $\beta_{\mathrm{2d}}$ of the JT black hole is given by
\begin{equation}\label{legame_beta}
\beta_{\mathrm{2d}}=\frac{4 G_3 C}{ \ell_{\text{AdS}}^2 } \beta_{\mathrm{3d}} \;,
\end{equation}
in terms of the inverse temperature $\beta_{\mathrm{3d}}$ of the BTZ. In \eqref{legame_beta} $\ell_{\text{AdS}}$ represents the AdS radius, $G_{\text{3d}}$ is the 3d Newton's constant, while $C$ is the usual coupling of JT gravity\footnote{
	In the JT literature, $C=\frac{\phi_r}{8 \pi G_2}$, where $\phi_r$ is the renormalized value of the dilaton on the boundary and $G_2$ the 2d Newton's constant.} \cite{Stanford:2017thb,Mertens:2022irh}.
This relation, which will turn useful later, can be proven by equating the corresponding entropies of the BTZ and JT black holes.\footnote{
	The entropy of the BTZ black hole is given by the Hawking formula $S_{\mathrm{BTZ}}=\frac{2 \pi r_{\text{h}}}{4G_{3}}$, where the shorthand $r_{\text{h}}=2\pi \ell_{\text{AdS}}^2/\beta_{\mathrm{3d}}$ is used to denote the radius of the event horizon \cite{Banados:1992wn}. The entropy of the JT black hole is instead given by $S_{\mathrm{JT}}=4 \pi^2 C/\beta_{\mathrm{2d}}$ \cite{Mertens:2022irh}. We expect $S_{\mathrm{BTZ}}=S_{\mathrm{JT}}$ in the strict s-wave reduction, connecting the three dimensional and the two-dimensional theory.}
However, as noticed in \cite{Mertens:2022ujr}, in most of the literature the JT gravity action is found to be supported in the $\left(r,\phi\right)$ section \cite{Jensen:2016pah,Maldacena:2016upp,Engelsoy:2016xyb}. For instance, in \cite{Cotler:2018zff} the Schwarzian model, which describes the boundary degree of freedom of JT gravity, emerges on the spatial angular direction, after compactifying the time circle. It turns out the correct identification for our purposes is this second one, where the coordinate $\varphi$, which we introduced in \eqref{EQ:hemisphere_metric} to parametrize the hemisphere, is to be interpreted as a spatial direction.
 
Based on the digression above, we now present a brief argument to identify the correct values of the unspecified scales $\ell$ and $L$ appearing in \eqref{EQ:S_tot}. Our starting point is the gauge formulation of 3d gravity with negative cosmological constant, which can be rephrased as a double Chern--Simons theory \cite{Witten:1988hc,Achucarro:1986uwr}:
\begin{equation}
	S = i (S_{\text{CS}}[A]-S_{\text{CS}}[\bar A]) \;,
\end{equation}
with action
\begin{equation}\label{chern}
S_{\textup{CS}}[A]=\frac{k}{4\pi}\int \Tr(A\wedge \mathrm{d}A+\frac{2}{3}A\wedge A\wedge A)\,.
\end{equation}
Here, $A$ and $\bar A$ are independent $\mathfrak{sl}(2)$ gauge fields, and $k$ is the Chern--Simons level, related to the gravitational parameters by $k=\frac{\ell_{\text{AdS}}}{4G_{3}}$. 
For our derivation, we focus on the holomorphic sector of the theory, which is described by the connection $A$. By virtue of what was argued before, we consider the Chern--Simons theory to be supported on a solid torus $D \times S^{1}$ which has the same topology of thermal AdS$_3$, with the Euclidean time coordinate running along the non-contractible cycle $S^{1}$. We equip it with the following metric:
\begin{equation}\label{tor}
\mathrm{d}s^2_{\mathrm{EAdS_3}} = \mathrm{d}t_{\mathrm{E}}^2 + C^2 \left(\text{d}\theta^2 +\sin^2\!\theta \, \mathrm{d}\varphi^2\right) \;,
\end{equation}
where the time variable is identified as $t_{\mathrm{E}}\sim t_{\mathrm{E}}+\beta_{\mathrm{2d}}k$.\footnote{One can easily prove that the metric \eqref{tor}, by using the identification $\beta_{\mathrm{2d}} = \frac{C}{k \ell_{\text{AdS}} } \beta_{\mathrm{3d}}$ given by \eqref{legame_beta} and by performing the change of variables $\theta=\arctan (r/\ell_{\text{AdS}})$, is Weyl equivalent to the metric
\[\mathrm{d}s^2_{\mathrm{EAdS_3}} = \left(\frac{r^2}{\ell_{\text{AdS}}^2}+1 \right)\mathrm{d}t_{\mathrm{E}}^2+\left(\frac{r^2}{\ell_{\text{AdS}}^2}+1\right)^{-1}\mathrm{d}r^2+r^2 \mathrm{d}\phi^2\]
of thermal AdS$_3$.} We then dimensionally reduce Chern--Simons theory \eqref{chern} along $t_{E}$ and by setting $\sigma=k^2 A_{t_{\mathrm{E}}}$ we obtain:
\begin{equation}
S = -i\frac{\beta_{\mathrm{2d}}}{2 \pi}\left(\int \Tr \left(\sigma F\right)-\frac{1}{2} \oint \mathrm{d}\varphi \, \Tr (\sigma A_{\varphi})\right) \;.
\end{equation}
By enforcing the boundary condition
\begin{equation}\label{bc}
A_{\varphi}=-i  C\sigma \;,
\end{equation}
we finally arrive at the action
\begin{equation}\label{our2}
S = \frac{\beta_{\mathrm{2d}}}{2 \pi}\left(-i\int \Tr \left(\sigma F\right)+\frac{C}{2} \oint \mathrm{d}\varphi \, \Tr (\sigma^2)\right) \;.
\end{equation}
After integrating in the auxiliary fields of the $\mathcal{N}=(2,2)$ vector multiplet, we note that the action \eqref{our2} precisely corresponds to our supersymmetric theory \eqref{EQ:S_tot}. Through the comparison between them, we arrive at the identification of the physical scales in the following manner: 
\begin{align}
	L &\equiv \frac{\beta_{\mathrm{2d}}}{2\pi} \;, & \ell &\equiv C \;.
\end{align}
We also note that setting $\ell = C$ is consistent with the dimensional reduction of the metric \eqref{tor} to the metric \eqref{EQ:hemisphere_metric} of an hemisphere with a radius of $C$. Finally, the boundary condition \eqref{bc} is consistent with a well-defined variational principle for the action \eqref{EQ:S_tot}, established by the condition \eqref{var_principle}.

\section{Localization of the supersymmetric BF model}\label{localization}
The partition function of certain supersymmetric gauge theories can be evaluated using the supersymmetric localization technique. The method relies on the fact that if a theory possesses a fermionic symmetry $\delta_Q$, we can deform the action by adding a $\delta_Q$-exact term $t\delta_Q \mathcal V$ without altering the result of the path integral.

The proof of this property is trivial and goes as follows. One introduces an auxiliary quantity $Z(t)$
\begin{equation}\label{Z(t)}
	Z(t) = \int [D\Phi] \; e^{-S[\Phi]-t\delta_{Q} \mathcal{V}}
\end{equation}
By construction, $Z(0)$ gives the initial partition function. If we assume that $\delta_{Q}^2 \mathcal{V}=0$,\footnote{In general, $\delta^2_Q$ does not vanish, but it yields a bosonic symmetry of the theory. Thus, requiring $\delta^2_Q \mathcal{V}=0$ is equivalent to the invariance of $\mathcal{V}$ under this bosonic symmetry.} we can easily show that\footnote{The validity of \eqref{Z(t)1} also assumes that the convergence of the path integral does not depend on $t$ and the measure of integration is invariant under $\delta_Q$.}
\begin{equation}\label{Z(t)1}
	\frac{\mathrm{d}}{\mathrm{d}t} \, Z(t) = \int [D\Phi] \; \delta_{Q}\!\left(\mathcal{V}e^{-S[\Phi]-t\delta_{Q} \mathcal V}\right),
\end{equation}
which, in turn, vanishes since it is the integral of $\delta_{Q}$-exact expression. Therefore,  $Z(t)$ is independent of $t$. We can evaluate the partition function computing \eqref{Z(t)} for any value of $t$. In particular, we can take the limit $t\to\infty$. If the bosonic part of $\delta_Q \mathcal{V}$ is positive-definite, the path integral is \emph{exactly} captured by the semiclassical expansion of $Z(t)$ around the saddle points of $\delta_Q \mathcal V$, rather than around those of the classical action. 

In the following, we will apply this technique to the supersymmetric BF theory \eqref{EQ:S_tot}. By virtue of the considerations of Section \ref{scales}, this procedure gives an alternative way to calculate the partition function of JT gravity.

\subsection{Localizing term}
Following \cite{Honda:2013uca}, we choose the localizing supercharge $\delta_Q = \delta_{\epsilon,\bar{\epsilon}}$, where the specific form of the Killing spinors $\epsilon$ and $\bar\epsilon$ is given in \eqref{preservedsusy}.
The localizing term reads
\begin{equation}\label{deltaV}
	\mathcal{V} \equiv t \, \delta_{0,\bar\epsilon}\int_{{\mathrm{HS}^2} } \mathrm{d}^2x \, \sqrt{g} \, \Tr(\frac{1}{2}\bar\lambda\gamma_3\lambda-2i D\sigma+i\eta^2) \;.
\end{equation}
The variation in \eqref{deltaV} will yield a true bulk term containing the Yang-Mills action,
\begin{align}\label{sym}
	\delta_Q \mathcal{V}_{\textup{bulk}}
	= \int_{\mathrm{HS}^2}\mathrm{d}^2x \; \sqrt{g} \; \frac{\bar\epsilon\epsilon}{2} \, \Tr \Big[ &\left(f+\frac{\eta}{\ell}\right)^2 + \mathrm{D}^\mu\eta\mathrm{D}_\mu\eta + \mathrm{D}^\mu\sigma\mathrm{D}_\mu\sigma - [\eta,\sigma]^2 + D^2 \cr
	& -\frac{i}{2}\left(\mathrm{D}_\mu\bar\lambda\gamma^\mu\lambda - \bar\lambda\gamma^\mu\mathrm{D}_\mu\lambda\right) + i\bar\lambda[\eta,\lambda]+ \bar\lambda\gamma^3[\sigma,\lambda] \Big] \;,
\end{align}
and a total divergence 
\begin{align}
	\delta_Q \mathcal{V}_\textup{bdry}
	&=\int_{\mathrm{HS}^2} \mathrm{d}^2x \, \sqrt{g} \, \mathrm{D}_\mu \Big(\Tr\Big[-i\bar\epsilon\gamma^\mu\gamma^3\epsilon \left(f + i[\eta,\sigma] + \frac{\eta}{\ell}\right)\sigma - i\bar\epsilon\gamma^\mu\gamma^\nu\epsilon\sigma \mathrm{D}_\nu\eta \cr
	&\qquad -i\epsilon_{\lambda\nu}\bar\epsilon\gamma^\mu\gamma^\lambda\epsilon\sigma \mathrm{D}^\nu\sigma + \epsilon^{\mu\nu}\eta\mathrm{D}_\nu\sigma + \bar\epsilon\gamma^\mu\epsilon D\sigma + \frac{i}{2}(\bar\lambda\gamma^3\epsilon)(\bar\epsilon\gamma^\mu\lambda) - \frac{i\bar\epsilon\epsilon}{4}\bar\lambda\gamma^\mu\lambda \Big]\Big) \;. \cr
\end{align}
With the help of Stokes' theorem, the latter translates into a  family of boundary contributions. We split the boundary terms into bosonic and fermionic ones.
For the bosonic part, after algebraic manipulations, we find
\begin{align}\label{bound1}
	\delta_Q \mathcal{V}_\textup{bos}
	& = 2\pi\ell\oint_{\partial \mathrm{HS}^2} \mathrm{d}\varphi \; \Tr\Big(-\ell\sigma \mathrm{D}^\theta \sigma + i\sigma f + \frac{i}{\ell}\eta\sigma \Big) \;,
\end{align}
where the integral is taken over the boundary circle of the hemisphere.
In \eqref{bound1}, we neglected all the terms proportional to $\bar\epsilon\epsilon=\cos\theta$, since their coefficient vanishes on the boundary. Moreover, in moving from the first to the second equality, we have dropped the first and third terms since they combine into a total derivative in $\varphi$ once we use that $\bar\epsilon\gamma^3\epsilon=1$. 

To establish the appropriate boundary conditions for the $t$-deformed theory, we consider the variational principle of the total action, which consists of the classical action $S_{\mathrm{new}}$ and the localizing term $\delta_Q \mathcal{V}$. Since $S_{\mathrm{new}}$ does not generate any boundary term for the fermions, we focus on the bosonic part of the variational principle, resulting in:
\begin{equation}
\begin{split}
	\delta_Q S_{\mathrm{bos}}
	&= \delta_Q S_{\mathrm{tot}} + t\,\delta_Q\mathcal{V}_{\textup{bulk}} + t\,\delta_Q\mathcal{V}_\textup{bdry} \\
	&= L\oint_{\partial \mathrm{HS}^2} \mathrm{d}\varphi \; \Tr\Big[-i \sigma \, \delta_Q\Big(A_{\varphi}+i \ell\sigma-\frac{t}{L}\Big(f+\frac{\eta}{\ell}+i \ell \mathrm{D}^{\theta}\sigma\Big)\Big) \\
	&\kern 10em + \mathrm{D}^{\theta}\eta\,\delta_Q\eta + \frac{t}{L}\Big(f+\frac{\eta}{\ell}\Big) \, \delta_Q(A_{\varphi}+i\ell\sigma)\Big] \;.
\end{split}
\end{equation}
Since $\eta$ already vanishes at the boundary due to \eqref{eta_bc}, in order to ensure a well-defined variational principle for any $t>0$, we need the following set of boundary conditions:
\begin{align}\label{interpolating_bc}
	F\big|_{\partial\mathrm{HS}^2} &= 0 \;, &
	\left(A_{\varphi}+i\ell\sigma-i t \frac{\ell}{L} D^{\theta}\sigma\right)\bigg|_{\partial\mathrm{HS}^2} &= c \;,
\end{align}
where $c$ is a constant.
It is worth noting that as $t\rightarrow 0$, the first condition in \eqref{interpolating_bc} could be relaxed (as it was in our previous discussion) while the second condition precisely reproduces the boundary condition introduced in \eqref{var_principle}. Furthermore, consistency with the gravitational condition \eqref{bc} requires us to set $c=0$ in this case as well.
In the opposite limit as $t\rightarrow \infty$, we observe that the second condition in \eqref{interpolating_bc} transforms into $D^{\theta}\sigma=0$.
Hence, the boundary conditions \eqref{interpolating_bc} effectively interpolate between the classical picture at $t=0$ and the limit $t\rightarrow \infty$, which will be the regime of interest in the localization process.

The fermionic part, on the other hand, can be massaged into
\begin{align}\label{ferm_boundary}
	\delta_Q \mathcal{V}_\textup{ferm}
	& = 2\pi\ell\oint_{\partial \mathrm{HS}^2} \mathrm{d}\varphi \; \Tr \Big(-\frac{i}{4}\bar\lambda\lambda \Big) \;.
\end{align}
A set of sufficient conditions to make the fermionic boundary term \eqref{ferm_boundary} vanish in our convention is given by $\lambda_1 = \lambda_2$ and $\bar\lambda_1 = \bar\lambda_2$ on $\partial\mathrm{HS}^2$.
This requirement nicely complements the boundary conditions \eqref{eta_bc} and \eqref{interpolating_bc} imposed on the bosonic sector. In the following, we will be interested in the limit $t\rightarrow \infty$, where they are given by:
\begin{align}
	\eta\big|_{\partial\mathrm{HS}^2} &= 0 \;, &
	F\big|_{\partial\mathrm{HS}^2} &= 0 \;, &
	D^{\theta}\sigma\big|_{\partial\mathrm{HS}^2} &= 0 \;.
\end{align}
This set of conditions has the advantage of being (manifestly) gauge invariant. However, we find using a gauge-fixed form of them more convenient. In particular, we 
use our gauge freedom to set $A_{\theta}=0$ on the boundary.  This choice makes our localization computation easier since we can exploit some results already present in the literature \cite{Honda:2013uca}. Then, the bosonic boundary conditions reduce to the following standard form:
\begin{align}\label{bos_boundary}
	\partial_{\theta}A_{\varphi} \big|_{\partial\mathrm{HS}^2} &= 0 \;, &
	\partial_{\theta}\sigma \big|_{\partial\mathrm{HS}^2} &= 0 \;, &
	\eta \big|_{\partial\mathrm{HS}^2} &= 0 \;, &
	A_{\theta} \big|_{\partial\mathrm{HS}^2} &= 0 \;,
\end{align}
which amounts to considering Dirichlet boundary conditions for $\eta$ and $A_{\theta}$, while instead using Neumann boundary conditions for $A_{\varphi}$ and $\sigma$.

A few comments are now in order.
Performing a path integral over bosonic fields involves a choice of a half-dimensional integration contour in the space of complex fields.
In particular, the integration contour for bosonic fields in Euclidean gauge theories must be chosen in such a way that the resulting gauge group is some compact subgroup of the complexification $G_\mathbb{C}$ of the original gauge group $G$, and the bosonic action is positive definite.
In supersymmetric localization this applies both to the original and the localizing actions separately.

In order to give meaning to the path integral of the supersymmetric BF theory at hand, defined at the classical level over the gauge group $\mathrm{SL}(2,\mathbb{R})$, we we find that the correct choice to reproduce the Schwarzian result is to pick a contour in such way that all fields are real and the resulting gauge group is $\mathrm{SU}(2)$, a compact subgroup of $\mathrm{SL}(2,\mathbb{C})$.

For a detailed discussion on the choice of integration contour for supersymmetric $\mathcal{N}=(2,2)$ gauge theories on the hemisphere we refer the reader to \cite{Hori:2013ika}.

\subsection{Localization locus}\label{moduli}
The minimum of the bosonic gauge sector in \eqref{sym} is realized when the following set of conditions holds
\begin{align}\label{locus}
	f &= -\eta \;, &
	\mathrm{D}_{\mu}\sigma = \mathrm{D}_{\mu}\eta &= 0 \;, &
	D &= 0 \;, &
	\left[\sigma,\eta\right] &= 0 \;.
\end{align}
The localization locus defined by \eqref{locus} is easy to characterize. Consider first the scalar field $\eta$.
Since it vanishes at the boundary because of \eqref{bos_boundary} and is covariantly constant ($\mathrm{D}_{\mu}\eta=0$), it must vanish everywhere.
This implies $F_{12}=0$. In other words, the gauge field $A$ must be a flat connection. However, since the hemisphere is contractible,  every flat connection is gauge equivalent to $A_{\mu}=0$. The only non-vanishing field is the scalar field $\sigma$, fixed to be an arbitrary constant $\sigma_0$. In summary, the  set of field configurations that satisfies \eqref{locus}, compatible with the boundary conditions \eqref{bos_boundary}, is  
\begin{align}\label{locus1}
	\sigma &= \frac{\sigma_0}{\ell} \;, &
	A_{\mu} &= 0 \;, &
	\eta &= 0 \;, &
	D &= 0 \;.
\end{align}
Similarly to the case of the Chern--Simons reformulation of three-dimensional gravity, the BF theory is (classically) equivalent to JT gravity only when the zweibein is invertible. At the perturbative level, this requirement is implemented by expanding the path integral around the geometrical (semiclassical) saddle point $e^a_\mu=\delta^a_\mu$ and $\omega=0$, which is quite far from the non-geometrical saddle obtained in \eqref{locus1}. When we turn on the localizing parameter $t$, we are instead allowing for the emergence of new saddle points that may compete and eventually replace the semiclassical one. We implicitly assume that complexification permits to deform the original semiclassical contour into a new one, picking the dominant contribution from the non-geometrical saddle $A_{\mu}=0$ (see \eqref{locus1}).

We now evaluate the total classical action \eqref{EQ:S_tot} on the locus \eqref{locus1}. The bulk term vanishes and we are left only with the boundary term
\begin{equation}
\left.S_{\mathrm{tot}}\right|_{\mathrm{locus}}=\frac{L}{2\ell}\oint_{\partial\mathrm{HS}^2
} \mathrm{d}\varphi \ \mathrm{Tr}\left(\sigma_0^2\right).
\end{equation} 
Therefore the infinite-dimensional path integral, which evaluates the partition function of JT gravity, localizes to a matrix model with the following structure:
\begin{equation}\label{integrale}
	\mathrm{Z}_{\mathrm{JT}} = \int_{\mathfrak{g}} \mathrm{d}\sigma_{0} \; \exp\left(-\frac{L}{2\ell}\oint_{\partial\mathrm{HS}^2} \mathrm{d}\varphi \; \Tr\left(\sigma_0^2\right)\right) \, \mathcal{Z}_{\text{1-loop}}\left[\sigma_0\right] \;.
\end{equation} 
In \eqref{integrale} $\mathcal{Z}_{\text{1-loop}}\left[\sigma_0\right]$ encodes the contributions of the one-loop determinants arising from the Gaussian integrals originating from the localizing term $\delta\mathcal{V}$ when we expand around the locus \eqref{locus1}. The subscript  $\mathfrak{g}$ on the integral means we are integrating over the Lie algebra of the gauge group.

Since the initial action is gauge invariant, the integrand in \eqref{integrale} turns out to be invariant under the adjoint action of the gauge group. We can use this freedom to diagonalize the matrix $\sigma_0$ through a gauge transformation and reduce the integral over the entire Lie algebra $\mathfrak{g}$ to an integral over a chosen realization of the Cartan subalgebra $\mathfrak{t}$. The Jacobian of this transformation will produce the usual Vandermonde determinant at the level of the integration measure. These steps are summarized by the following general identity that holds for any integral of an adjoint invariant function $\mathfrak{f}(\sigma)$ 
\begin{equation}\label{integrale2}
	\frac{1}{\mathrm{vol}(\mathfrak{g})} \int_{\mathfrak{g}} \mathrm{d}^{d_{g}}\sigma \; \mathfrak{f}(\sigma) = \frac{1}{|W|} \int_{\mathfrak{t}} \mathrm{d}^{l_{\mathfrak{g}}}\sigma \; \mathfrak{f}(\sigma) \, \prod_{\alpha \in \Delta_{+}} \alpha(\sigma)^2 \;.
\end{equation} 
Above $d_{\mathfrak{g}}$ and $l_{\mathfrak{g}}$ are  the dimension and the rank of $\mathfrak{g}$, while $\Delta_{+}$ is the set of its positive roots, denoted with $\alpha$.  In \eqref{integrale2}, we normalize the l.h.s. by the order of the Weyl group, $\left|W\right|$, to account for the residual gauge symmetry. Then we are left with
\begin{equation}\label{integrale3}
	\mathrm{Z}_{\mathrm{JT}} = \frac{1}{2!} \int_{\mathfrak{t}} \mathrm{d}\sigma_{0} \; \alpha(\sigma_0)^2 \, \exp\left(-\frac{L}{2\ell}\oint_{\partial\mathrm{HS}^2} \mathrm{d}\varphi \; \Tr(\sigma_0^2)\right) \, \mathcal{Z}_{\text{1-loop}}[\sigma_0]
\end{equation} 
where $\sigma_0$ is now assumed to be in the Cartan, i.e., the component along the diagonal generator $\gamma_3$.

\subsection{One-loop determinants}\label{1l}
We now turn to the detailed evaluation of the one-loop determinants producing $\mathcal{Z}_{\text{1-loop}}$.
The analysis of the possible contributions for the case of  $\mathcal{N}=(2,2)$ theories on the hemisphere was done in detail in \cite{Sugishita:2013jca, Honda:2013uca, Hori:2013ika}. Below, we shall simply review the essential steps of the calculation and collect the
relevant results.

To begin with, we expand  each field of our supersymmetric model around the background value given by the localization locus \eqref{locus1}. Schematically, $\Phi \mapsto \Phi_0 + \hat\Phi/\sqrt{t}$.

Plugging this into the localizing (bulk) term \eqref{sym} and subsequently expanding in $t$, we can easily single out quadratic part of $\delta_Q\mathcal{V}$. Since this quantity vanishes on the locus \eqref{locus1}, we do not have any ``classical contribution", and we can write
\begin{align}\label{locas}
	\delta_Q \mathcal{V}^{(2)}
	=&\lim_{t\rightarrow \infty} \ t\int_{\mathrm{HS}^2} \delta_Q \mathcal{V} \cr
	=&\int_{\mathrm{HS}^2} \mathrm{d}^2 x \sqrt{\mathrm{g}}\ \mathrm{Tr} \left[-\hat{A}_{\mu} \nabla_{\nu}\nabla^{\nu} \hat{A}^{\mu}+ \hat{A}_{\mu} \nabla_{\nu}\nabla^{\mu} \hat{A}^{\nu}+\frac2\ell\hat{\eta}\ \epsilon^{\mu \nu}\nabla_{\mu}\hat{A}_{\nu}+\frac{\eta^2}{\ell^2}\right. \cr
	&-\frac{1}{\ell^2}[\sigma_0,\hat{A}^{\mu}][\sigma_0,\hat{A}_{\mu}]- \hat{\sigma}\nabla_{\mu}\nabla^{\mu}\hat{\sigma}-\frac{2i}{\ell}[\sigma_0,\hat{A}^{\mu}]\nabla_{\mu}\hat{\sigma}-\hat{\eta} \nabla_{\mu}\nabla^{\mu }\hat{\eta} \cr
	&+\hat{D}^2-\frac{1}{\ell^2}\left[\sigma_0,\hat{\eta}\right]^2 \left.+\frac{i}{2}\hat{\bar\lambda} \gamma^{\mu}\nabla_{\mu}\hat{\lambda}-\frac{i}{2}\nabla_\mu\hat{\bar{\lambda}} \gamma^{\mu}\hat{\lambda}+\frac{1}{\ell}\hat{\bar{\lambda}}\gamma_3[\sigma_0,\hat{\lambda}]\right] \;,
\end{align}
where we have integrated by parts some terms taking advantage of the boundary conditions for the fluctuation fields.

The quadratic integral over $D$ is trivial; thus, we can neglect it in the localizing term. From now on, we will omit the hat used to denote the fluctuation fields since this does not cause any ambiguity and allows for simpler notation.

\paragraph{Gauge fixing.}
Before moving on, we must gauge-fix the theory \cite{Pestun:2007rz} to remove the gauge redundancy. We choose  to impose the Lorentz gauge and set:
\begin{equation}
	\nabla_{\mu}A^{\mu}=0.
\end{equation}
To do so, we exploit the standard BRST construction by introducing two ghost fields $c$, $\bar{c}$, and a Lagrange multiplier $b$, all living in the adjoint representation of the gauge algebra. Next, we add the following term\footnote{The total susy transformation would now become $\delta_Q \mapsto \delta_Q + \delta_{\mathrm{BRST}}$.} to the localizing action \eqref{sym}
\begin{equation}\label{ghost}
	t\int_{\mathrm{HS}^2} \delta_Q \mathcal{V}_{\mathrm{brst}}=t\int_{\mathrm{HS}^2} \mathrm{d}^2 x \sqrt{\mathrm{g}} \ \mathrm{Tr}\left(\bar{c} \ \nabla_{\mu}\mathcal{D}^{\mu}c+b \ \nabla_{\mu}A^{\mu} \right).
\end{equation}
Expanding it as before around the combined  locus \eqref{locus1} and $c=\bar{c}=b=0$,  we get
\begin{equation}\label{ghost2}
	\delta_Q \mathcal{V}_{\mathrm{brst}}^{(2)}=\int_{\mathrm{HS}^2} \mathrm{d}^2 x \sqrt{\mathrm{g}} \ \mathrm{Tr}\left(\frac12 \bar{c} \ \nabla_{\mu}\nabla^{\mu}c+b \ \nabla_{\mu}A^{\mu} \right).
\end{equation}
The integration over the bosonic Lagrange multiplier $b$ gives $\delta(\nabla_{\mu}A^{\mu})$, which enforces the gauge-fixing condition in the path integral.

Following \cite{Kapustin:2009kz}, we then separate the gauge field into a divergenceless and pure divergence part
\begin{equation}
	A_{\mu}
	= \partial_{\mu}u + A_{\mu}' \;,
\end{equation}
where $A_{\mu}'$ is the divergenceless part of $A_{\mu}$, i.e.\ $\nabla_{\mu}A'{}^{\mu} = 0$. Exploiting this decomposition, the delta function imposing the gauge-fixing becomes $\delta\left(-\nabla^2 u\right)$, with $\nabla^2\equiv\nabla^\mu\nabla_\mu$. Thus, the integration measure for the gauge field can be  rewritten as follows
\begin{align}
	[DA_{\mu}] \; \delta(\nabla^{\mu}A_{\mu})
	&= [DA'_{\mu}] [Du] \; \delta(-\nabla^2 u) \cr
	&= [DA'_{\mu}][Du] \; \delta(u)\det(-\nabla^2)^{-\frac{1}{2}} \;.
\end{align}
The scalar $u$ can then be integrated out, leaving only the Jacobian factor $\det(-\nabla^2)^{-1/2}$. Subsequently, we can perform the functional integrations over $\sigma$ and the ghosts. The former gives an additional factor $\det(-\nabla^2)^{-1/2}$, while the latter provides a factor of $\det(-\nabla^2)$, so that the above three contributions exactly cancel. 

The gauge fixed quadratic localizing term now reads 
\begin{align}\label{locas3}
	\delta_Q \mathcal{V}^{(2)}_{\text{g.f.}}
	=& \int_{\mathrm{HS}^2} \mathrm{d}^2 x \sqrt{\mathrm{g}} \ \mathrm{Tr} \bigg[- A_{\mu}' \nabla_{\nu}\nabla^{\nu} A_{\mu}'+\frac{1}{\ell^2}A_{\mu}'A^{'\mu}+\frac{2}{\ell}\eta \ \epsilon^{\mu \nu}\nabla_{\mu}A_{\nu}'- \frac{1}{\ell^2}[\sigma_0,A'^{\mu}][\sigma_0,A_{\mu}'] \cr
	&- \eta \nabla_{\mu}\nabla^{\mu }\eta+\eta^2-\frac{1}{\ell^2}[\sigma_0,\eta]^2 +\frac{i}{2}\lambda \gamma^{\mu}\nabla_{\mu}\bar{\lambda}+\frac{i}{2}\bar{\lambda} \gamma^{\mu}\nabla_{\mu}\lambda+\frac{1}{\ell}\bar{\lambda}\gamma_3[\sigma_0,\lambda]\bigg] \;.
\end{align}
The next step consists in using the Cartan decomposition, that is, to expand the adjoint field $X$ as 
\begin{equation}
	X = \sum_{i} X^{i} \mathsf{H}_{i} + \sum_{\alpha \in \Delta_{+}} (X^{\alpha}\mathsf{E}_{\alpha}+X^{-\alpha}\mathsf{E}_{-\alpha}) \;,
\end{equation}
where $\mathsf{H}_{i}$ are the Cartan generators, $\mathsf{E}^{\alpha}$ is the generator corresponding to the root $\alpha$ and $\Delta_{+}$ is the set of positive roots. They satisfy the following relations
\begin{align}\label{relations}
	[\mathsf{H}_{i},\mathsf{E}_{\alpha}] &= \alpha(\mathsf{H}_i)\mathsf{E}_{\alpha} \;, &
	\mathsf{E}_{\alpha}^{\dagger} &= \mathsf{E}_{-\alpha} \;, &
	\Tr\left(\mathsf{E}_{\alpha}\mathsf{E}_{\beta}\right) &= \delta_{\alpha+\beta} \;, &
	\Tr\left(\mathsf{E}_{\alpha}\mathsf{H}_{i}\right) &= 0 \;.
\end{align}
For $\mathfrak{sl}(2,\mathbb{R})$, we only have one Cartan generator $\gamma_3$ and one positive root $\alpha$. Therefore, we shall drop the sum over the positive roots in the following.

\paragraph{Bosonic determinants.}
Using the commutation and trace relations \eqref{relations} one can find the bosonic part of \eqref{locas3} is proportional to
\begin{align}\label{locas4}
\delta_Q \mathcal{V}^{(2)}_{\mathrm{bos}}=&\int_{\mathrm{HS}^2} \mathrm{d}^2 x \sqrt{\mathrm{g}} \ \left[- A_{\mu}^{-\alpha } \nabla_{\nu}\nabla^{\nu} A_{\mu}^{\alpha }+ \frac{1}{\ell^2}A_{\mu}^{-\alpha}A^{\alpha,\mu}+ \frac{1}{\ell}\eta^{-\alpha} \ \epsilon^{\mu \nu}\nabla_{\mu}A_{\nu}^{\alpha}+ \frac{1}{\ell}\eta^{\alpha} \ \epsilon^{\mu \nu}\nabla_{\mu}A_{\nu}^{-\alpha}\right. \nonumber
\\
&+\frac{1}{\ell^2}\alpha (\sigma_{0})^2 A_{\mu}^{-\alpha }A^{\alpha \mu }\left.- \eta^{-\alpha} \nabla_{\mu}\nabla^{\mu }\eta^{\alpha}+\frac{1}{\ell^2}\eta^{-\alpha}\eta^{\alpha}+\frac{1}{\ell^2}\alpha (\sigma_{0})^2 \eta^{-\alpha }\eta^{\alpha }
\right],
\end{align}
where we omitted the prime and implicitly considered only the divergenceless part of $A_{\mu}$.
We find it convenient to expand the gauge field in terms of the vector spherical harmonics $\mathcal{C}^{\lambda}_{jm,\mu}$ and to write:
\begin{equation}\label{generalexpansion}
	A_{\mu}^{\alpha} = \sum_{\lambda=1,2} \sum_{j=1}^{\infty}\sum_{m=-j}^{j} \ A^{\alpha,\lambda}_{jm} \, \mathcal{C}^{\lambda}_{jm,\mu}(\vartheta,\varphi) 
\end{equation}
These special functions enjoy the following two properties
\begin{align}
	\nabla^{\mu}\mathcal{C}_{jm,\mu}^{1} &= -\frac{\sqrt{j(j+1)}}{\ell^2} \, \mathcal{Y}_{jm} \;, &
	\nabla^{\mu}\mathcal{C}_{jm,\mu}^{2} &= 0 \;.
\end{align}
We indicated the usual scalar spherical harmonics with $\mathcal{Y}_{jm}$. Since $A_\mu$ is divergenceless, only the component with helicity $\lambda=2$ can appear in the above expansion. Therefore, we can drop the sum over $\lambda$ in \eqref{generalexpansion} and write 
\begin{align}\label{lambda2}
	A_{\mu}^{\alpha} &= \sum_{j=1}^{\infty} \sum_{m=-j}^{j} A^{\alpha,2}_{jm} \, \mathcal{C}^{2\vphantom{,}}_{jm,\mu}(\vartheta,\varphi) \;, &
	A_{\mu}^{-\alpha} &= (A_{\mu}^{\alpha})^{*} \;.
\end{align}
The boundary conditions satisfied by $A_\mu$ further restrict this sum, and the coefficients $A^{\alpha,2}_{jm}$ are different from zero only when $j-m$ is an odd integer (see \cite{Honda:2013uca}). Similarly, we can expand the scalar field $\eta$ in terms of the usual spherical harmonics:
\begin{align}\label{sigma2}
	\eta^{\alpha} &= \sum_{j=0}^{\infty} \sum_{m=-j}^{j}  \eta^{\alpha}_{jm} \, \mathcal{Y}^{\vphantom{\alpha}}_{jm}(\vartheta,\varphi) \;, &
	\eta^{-\alpha} &= \left(\eta^{\alpha}\right)^{*} \;.
\end{align}
The vanishing of this field at the boundary again imposes that the expansion coefficients differ from zero only when $j-m=\mathrm{odd}$ \cite{Honda:2013uca}.

Both scalar and vector spherical harmonics are eigenvectors of the corresponding Laplacian, though with different eigenvalues, i.e.
\begin{align}\label{eige}
	-\nabla^{\mu}\nabla_{\mu} \, \mathcal{Y}_{jm} &= \frac{j(j+1)}{\ell^2} \, \mathcal{Y}_{jm} \;, &
	-\nabla^{\mu}\nabla_{\mu} \, \mathcal{C}_{jm}^{\lambda} &= \frac{j(j+1)-1}{\ell^2} \, \mathcal{C}_{jm}^{\lambda} \;.
\end{align}
Moreover, the vector harmonics satisfy these further set of relations
\begin{equation}\label{rotor}
	\epsilon^{\mu \nu} \nabla_{\mu} \left(\mathcal{C}^{\lambda}_{jm}\right)_{\nu}=-\delta^\lambda_2\frac{\sqrt{j(j+1)}}{\ell^2}\mathcal{Y}_{jm}.
\end{equation}
The above properties will allow us to deal with the mixed terms present in the bosonic sector. Then, by taking advantage of \eqref{eige} and \eqref{rotor}  as well as  the  orthogonality relations on the hemisphere
\begin{align}\label{ortogo}
	\int_{0}^{2\pi} \mathrm{d} \varphi \int_{0}^{\frac{\pi}{2}} \mathrm{d}\vartheta \; \sin\vartheta \; \mathcal{Y}_{jm}(\vartheta,\varphi)^{*} \; \mathcal{Y}_{j'm'}(\vartheta,\varphi) &= \frac12\delta_{jj'} \delta_{mm'} \;, \\
	\int_{0}^{2\pi} \mathrm{d} \varphi \int_{0}^{\frac{\pi}{2}} \mathrm{d}\vartheta \; \sin\vartheta \; \mathcal{C}_{jm,\mu}^{\lambda}(\vartheta,\varphi)^{*} \; \mathcal{C}_{j'm',\nu}^{\lambda'}(\vartheta,\varphi) \; g^{\mu\nu} &= \frac12\delta_{jj'} \delta_{mm'} \delta^{\lambda \lambda'} \;,
\end{align}
we can easily show that the bosonic sector of $\delta\mathcal{V}^{(2)}$  \eqref{locas4} reduces to
\begin{equation}\label{ma3}
\delta_Q \mathcal{V}^{(2)}_{\mathrm{bos}}=\sum_{j,m} \left(\Q^{\alpha}_{jm}\right)^{\dagger}\Delta^{\mathrm{bos}}_{j}  \left(\Q^{\alpha}_{jm}\right)
\end{equation}
where  $\Q^{\alpha}_{jm}=\left(A^{\alpha, 2} \ \ \eta^{\alpha}\right)^{\mathrm{T}}_{jm}$. The explicit form of the matrix $\Delta^{\mathrm{bos}}_{j}$ is
\begin{equation}\label{ma2}
\Delta^{\mathrm{bos}}_{j}=\frac{1}{\ell^2}
\left(\begin{matrix} j\left(j+1\right)+\alpha(\sigma_0)^2 & \sqrt{j\left(j+1\right)} \\ \\ \sqrt{j\left(j+1\right)} & j\left(j+1\right)+\alpha(\sigma_0)^2+1 
\end{matrix}\right).
\end{equation}
Recall now that  $\Q^{\alpha}_{jm}$ vanishes when $j-m$ is even because of the boundary conditions. This fact reduces the usual degeneracy in $m$ of a spherical symmetric problem from $2j+1$ to $j$. Thus, the total bosonic contribution to $\mathcal{Z}_{\text{1-loop}}$ will read \cite{Honda:2013uca} (taking into account that integration variable $\Q^{\alpha}_{jm}$ is complex)
\begin{align}\label{bosonico}
	\mathcal{Z}_{\text{1-loop}}^{\mathrm{bos}}
	&= \prod_{j} \left(\det\Delta^{\mathrm{bos}}_{j}\right)^{-j} \cr
	&= \prod_{j=1}^{\infty} \frac{1}{\ell^{4j}[j^2+\alpha(\sigma_{0})^2]^{j} [(j+1)^2+\alpha(\sigma_{0})^2]^{j}} \;.
\end{align}

\paragraph{Fermionic determinant.}
The fermionic part of \eqref{locas3} after Cartan decomposition and some integration by parts becomes
\begin{equation}\label{matric}
	\delta_Q\mathcal{V}^{(2)}_{\mathrm{fer}} = \int \mathrm{d}^2x \; \sqrt{\mathrm{g}} \
	\left(\lambda^{\alpha} \ \bar{\lambda}^{\alpha}\right)^{\dagger}\left(\begin{matrix} 0 & i\gamma_3\gamma^{\mu}\nabla_{\mu}-\frac{1}{\ell}\alpha (\sigma_0) \\ i\gamma_3\gamma^{\mu}\nabla_{\mu}+\frac{1}{\ell}\alpha (\sigma_0) & 0 \end{matrix}\right) \left(\begin{matrix} \lambda^{\alpha} \\ \bar{\lambda}^{\alpha}\end{matrix}\right).
\end{equation}
The symbol $\dagger$ denotes a Dirac-like conjugation containing also a factor $\gamma_3$, that is 
\begin{equation}
\left(\lambda^{\alpha} \ \bar{\lambda}^{\alpha}\right)^{\dagger}=\left(\lambda^{-\alpha}\gamma_3\,, \ \bar{\lambda}^{-\alpha}\gamma_3\right).
\end{equation}
The gauginos $\lambda$, $\bar{\lambda}$ are fields of spin $1/2$  and we denote the two different helicities with  $s=\pm (\pm\frac12)$. We can expand both of  them into spin spherical harmonics $\mathcal{Y}^{s}_{jm}$:
\begin{align}\label{expaspin}
	\lambda^{\alpha}
	&= \sum_{s=\pm} \sum_{j=\frac12}^{\infty} \sideset{}{^{'}} \sum_{m=-j}^{j} \lambda^{\alpha,s}_{jm} \, \mathcal{Y}^{s\vphantom{,}}_{jm}(\vartheta,\varphi)  \;, &
	\bar{\lambda}^{\alpha}
	&= \sum_{s=\pm} \sum_{j=\frac12}^{\infty} \sideset{}{^{'}}\sum_{m=-j}^{j} \bar{\lambda}^{\alpha,s}_{jm} \, \mathcal{Y}^{s\vphantom{,}}_{jm}(\vartheta,\varphi) \;.
\end{align}
The spin spherical harmonics are eigenvectors of the Dirac operator, i.e.
\begin{equation}\label{Dirac}
	i \gamma_3\gamma^{\mu}\nabla_{\mu} \mathcal{Y}^{\pm}_{jm} = \pm\frac{i}{\ell} \left(j+\frac12\right) \, \mathcal{Y}^{\pm}_{jm}
\end{equation}
with $j=\frac12,\frac32,\cdots$, $m=-j,-j+1,\cdots,j,$ and they are normalized on the hemisphere with
\begin{equation}
	\int_{0}^{2\pi} \mathrm{d}\varphi \int_{0}^{\frac{\pi}{2}} \mathrm{d}\vartheta \; \sin (\vartheta) \, \mathcal{Y}_{jm}^{s}(\vartheta,\varphi)^{*} \; \mathcal{Y}_{j'm'}^{s'}(\vartheta,\varphi) = \frac{1}{2} \ \delta_{jj'} \delta_{mm'}\delta^{ss'} \;.
\end{equation}
In \eqref{same2},
the prime in the internal sum over $m$ means that we must restrict the values of $m$ to even $j-m$ if $s=+$ and to even $j-m$ if $s=-$. This constraint stems from the boundary conditions imposed on the fermions \cite{Honda:2013uca}.

Plugging the expansion \eqref{same2} into  \eqref{matric} and exploiting the orthogonality relations to perform the angular integrations, we find the fermionic term \eqref{matric} can be reorganized  into the sum of two series
\begin{align}\label{same2}
	\delta_Q\mathcal{V}^{(2)}_{\mathrm{fer}}
	&= \delta_Q\mathcal{V}^{(2)}_{\mathrm{fer,+}} + \delta_Q\mathcal{V}^{(2)}_{\mathrm{fer,-}} \cr
	&= \sum_{j=\frac12}^{\infty}\sideset{}{^{'}}\sum_{m=-j}^{j} (\Lambda^{\alpha,+}_{jm})^{\dagger} \ \Delta_{j}^{\mathrm{fer,+}} \ (\Lambda^{\alpha,+}_{jm}) + \sum_{j=\frac12}^{\infty}\sideset{}{^{'}}\sum_{m=-j}^{j}(\Lambda^{\alpha,-}_{jm})^{\dagger} \ \Delta_{j}^{\mathrm{fer,-}} \ (\Lambda^{\alpha,-}_{jm}) \;,
\end{align}
where $\Lambda^{\alpha,\pm}_{jm}=\left(\lambda^{\alpha,\pm} \  \ \bar{\lambda}^{\alpha,\pm}\right)^{\mathrm{T}}_{jm}$ with
\begin{equation}\label{same}
\Delta_{j}^{\mathrm{fer,\pm}}=\frac{1}{\ell}\left(\begin{matrix} 0 &\pm j\pm\frac12+i\alpha(\sigma_0) \\ \pm j\pm \frac12-i\alpha(\sigma_0) & 0 \end{matrix}\right) \;.
\end{equation}
Since this matrix depends only on $j$, we have the usual degeneracy in $m$ for each eigenvalue. This degeneracy is reduced from $2j+1$ to $j+1/2$ by the constraint on the sum over $m$.

Since the matrices $\Delta_{j}^{\mathrm{fer,+}}$ and $\Delta_{j}^{\mathrm{fer,-}}$ possess the same determinant,   the total fermionic contribution to $\mathcal{Z}_{\text{1-loop}}^{\mathrm{fer}}$ can be written, up to a phase, as follows \cite{Benini:2012ui}:
\begin{equation}\label{fermionic}
\begin{split}
	\mathcal{Z}_{\text{1-loop}}^{\mathrm{fer}}=&\prod_{j=\frac12}^{\infty} \left(\det \Delta_{j}^{\mathrm{fer,+}}\right)^{j+\frac12}\left(\det \Delta_{j}^{\mathrm{fer,-}}\right)^{j+\frac12} \\
	=&\prod_{j=\frac12}^{\infty} \ell^{4j+2}\left(j+\frac12 +i\alpha (\sigma_{0})\right)^{2j+1}\left(j+\frac12 -i\alpha (\sigma_{0})\right)^{2j+1}
\end{split}
\end{equation}

\paragraph{Total one-loop contribution.}
Collecting together the bosonic and fermionic contributions \eqref{bosonico} and \eqref{fermionic}
and simplifying the common factors between the numerator and denominator, we arrive at
\begin{align}
	\mathcal{Z}_{\text{1-loop}}
	&= \mathcal{Z}_{\text{1-loop}}^{\mathrm{bos}} \cr
	&\sim \prod_{j=1}^{\infty} \left(j^2+\alpha(\sigma_0)^2\right) \cr
	&\sim \bigg(\prod_{j=1}^{\infty}j^2\bigg) \prod_{j=1}^{\infty} \bigg(1+\frac{\alpha(\sigma_0)^2}{j^2}\bigg) \;.
\end{align}
The first of the two infinite products can be regularized by using the zeta function regularization, and we get
\begin{equation}
	\prod_{j=1}^{\infty}j^2
	= e^{-2\zeta'(0)}
	= 2\pi \;,
\end{equation}
where we used $\zeta'(0)=-\frac12 \ln 2\pi$.
In the second one, we recognize the representation of the hyperbolic sine as an infinite product. Then
\begin{equation}\label{one}
	\mathcal{Z}_{\text{1-loop}} = \frac{2\sinh(\pi \alpha (\sigma_0))}{\alpha(\sigma_0)} \;.
\end{equation}

\subsection{Result}\label{res}
By substituting the one-loop determinant \eqref{one} into the localization formula \eqref{integrale3}, we find that the partition function is given by
\begin{equation}\label{integrale4}
	\mathrm{Z}_{\mathrm{JT}} = \frac{1}{2} \int_{\mathfrak{t}} \mathrm{d}\sigma_{0} \; \alpha(\sigma_0)^2\exp\bigg({-}\frac{L}{2\ell} \oint_{\partial\mathrm{HS}^2} \mathrm{d}\varphi \; \Tr(\sigma_0^2)\bigg) \; \frac{2\sinh \pi \alpha (\sigma_0)}{\alpha(\sigma_0)} \;.
\end{equation}
We observe that the denominator of $\mathcal{Z}_{\text{1-loop}}$ cancels exactly with one factor of $\alpha(\sigma_0)$ arising from the Vandermonde determinant, resulting in
\begin{equation}\label{integrale5}
	\mathrm{Z}_{\mathrm{JT}} = \int_{\mathfrak{t}} \mathrm{d}\sigma_{0} \; \alpha(\sigma_0) \, \sinh(\pi \alpha (\sigma_0)) \, \exp\bigg({-}\frac{\pi L}{\ell}\Tr(\sigma_0^2)\bigg) \;.
\end{equation}
Since $\sigma_0$ lies in the Cartan subalgebra $\mathfrak{t}$, we can parameterize it as $\sigma_0 = s\gamma_3$ with $s \in \mathbb{R}$. For $\mathfrak{su}(2)$, the only positive root is $1$, so we have $\alpha(s \gamma_3)=2s$ and $\Tr(\gamma_3^2) = 2$. Consequently, \eqref{integrale5} becomes
\begin{equation}\label{integrale6}
	\mathrm{Z}_{\mathrm{JT}}= 4\int_{0}^{\infty} \mathrm{d}s \; s \, \sinh(2\pi s) \, \exp\bigg({-}\frac{\beta}{C} s^2\bigg) \;,
\end{equation}
where we have reinstated the gravitational scales $\ell=C$ and $\beta_{\mathrm{2d}} \equiv \beta = 2\pi L$, and utilized the integrand's parity to limit the integral to the range $[0,+\infty)$. This result \eqref{integrale6} coincides with the one obtained in \cite{Iliesiu:2019xuh} through $\mathrm{SL}(2,\mathbb{R})$ Hamiltonian quantization.

We can now evaluate the integral over $s$ and obtain
\begin{equation}\label{integrale8}
\mathrm{Z}_{\mathrm{JT}}\propto \left(\frac{C\pi}{\beta}\right)^{3/2} e^{\frac{C\pi^2}{\beta}},
\end{equation}
which reproduces the result obtained through equivariant localization of the Schwarzian theory  \cite{Stanford:2017thb} and  via conformal bootstrap from Liouville CFT \cite{Mertens:2017mtv}.

\section{Higher-spin JT gravity}\label{higher}
Our results can be easily generalized to compute the partition function of the higher-spin version of JT gravity on the disk.
Here, we are interested in higher-spin theories living on an AdS background.
The first remarkable constructions were developed in four dimensions \cite{Fradkin:1987ks, Vasiliev:1990en, Vasiliev:1992av, Vasiliev:1995dn, Vasiliev:1999ba}, and later extended to a generic number of spacetime dimensions \cite{Vasiliev:2003ev}. 
higher-spin theories also display an essential role in holography \cite{Klebanov:2002ja, Gaberdiel:2010pz}. 

One may hope that in lower dimensions, some simplifications happen. That is the case for three-dimensional higher-spin gravity, as there are no propagating local degrees of freedom. Moreover, higher-spin theories admit a Chern--Simons formulation \cite{Blencowe:1988gj} that generalizes the pure gravitational construction. Unlike the higher dimensional cases, there is no need to consider an infinite number of higher-spin fields for having consistent interactions \cite{Aragone:1983sz, Campoleoni:2010zq, Campoleoni:2011hg}. A natural and simple example is the higher-spin theory corresponding to the Chern--Simons theory $\mathrm{SL}(N, \mathbb{R})\times \mathrm{SL}(N,\mathbb{R})$, containing fields with spin up to $N$. 

A somewhat analogous situation also occurs in higher-spin extensions of JT gravity. They can be constructed from the gauge theory formulation and allowing the gauge group to be $\mathrm{SL}(N, \mathbb{R})$ \cite{Alkalaev:2013fsa,Grumiller:2013swa, Alkalaev:2014qpa}.\footnote{Here we are not precise on the global structure of the gauge group, as it will not be relevant.}
While all these formulations require some relevant modifications of the considerations worked out in the standard JT gravity, see for instance \cite{Gonzalez:2018enk, Narayan:2019ove, Datta:2021efl, Kruthoff:2022voq}, the technology developed in the previous chapter can be readily extended to the higher-spin case $\mathrm{SL}(N, \mathbb{R})$. 

\subsection{The supersymmetric higher-spin theory}

To begin with, we shall briefly review how an $\mathrm{SL}(N, \mathbb{R})$ version of the BF theory realizes a higher-spin theory \cite{Alkalaev:2013fsa}. As done for standard JT gravity in Section~\ref{sec1}, we work with the first-order formalism and organize fields into a connection and dilaton field. Let us now be slightly more general and work with a gauge group $G$ with the property that it contains an $\mathrm{SL}(2,\mathbb{R})$ sector generated by the $\mathsf{P}_{i}$ of Section~\ref{sec1}. This factor corresponds to the AdS$_2$ isometry group. Then we demand that all the other generators in the adjoint of $G$ can be decomposed into a totally symmetric irreducible representation of the $\mathrm{SL}(2,\mathbb{R})$ factor. Let us denote these generators as $\mathsf{T}_{a_1\dots a_s}$ for some integer $s$. They are totally symmetric and traceless, namely $\eta^{a_1a_2} \mathsf{T}_{a_1 a_2\dots a_s} = 0$.

Any Lie-algebra valued field $\Phi$ in our BF theory will have an expansion
\begin{equation}\label{high_spin_exp}
	\Phi = \Phi^a \, \mathsf{P}_a + \sum_s \Phi^{a_1\dots a_s} \, \mathsf{T}_{a_1\dots a_s}\,.
\end{equation}
Because of the specific properties of $\Phi^{a_1\dots a_s}$, it is natural to interpret it as a higher-spin $s$ field \cite{Campoleoni:2010zq, Campoleoni:2011hg}. 

It turns out that $\mathrm{SL}(N,\mathbb{R})$ satisfies all the above conditions \cite{Bergshoeff:1989ns}. Therefore, from now on, we will focus on this specific case and construct the corresponding generalization of JT gravity. To do so, we mimic the BF construction of Sec. \ref{sec1}. That is, we introduce an $\mathrm{SL}(N,\mathbb{R})$ dilaton field $\chi$ expanded as in \eqref{high_spin_exp} and a $\mathrm{SL}(N,\mathbb{R})$ connection
\begin{equation}
	A = \sum_s A_\mu^{a_1\dots a_s} \, \mathsf{T}_{a_1 a_2\dots a_s} \, \mathrm{d}x^\mu \;.
\end{equation}
The action is just the BF one
\begin{equation}
	S_{\mathrm{BF}} = -i \int_{\Sigma} \Tr \left(\chi F\right)
\end{equation}
The equation of motions are
\begin{align}
	F_{\mu\nu}^{a_1\dots a_s} &= 0 \;, &
	D_\mu \chi^{a_1\dots a_s} &= 0 \;,
\end{align}
where $\mathrm{D}_\mu=\nabla_\mu+A_\mu$ and $F_{\mu\nu}^{a_1\dots a_s}$ is the field strength related to $A_\mu^{a_1\dots a_s}$. One can study them in the metric formulation around the AdS$_2$ background and show that indeed reproduce a 2d higher-spin gravity theory, identifying the spectrum \cite{Alkalaev:2013fsa}.

The next step would be studying the asymptotic boundary conditions that reproduce a consistent generalization of the Schwarzian dynamics \cite{Gonzalez:2018enk, Kruthoff:2022voq}.
This requires the addition of the familiar boundary term
\begin{equation}
	S_{\textup{bdy}}\propto \oint_{\partial \Sigma} \mathrm{d}\varphi \Tr(\chi^2) \;.
\end{equation}
Not so surprisingly, one must also consider asymptotic boundary conditions preserving the $W_N$-algebra (a nonlinear extension of Virasoro).

However, we do not follow this approaches here. We rather adopt an extension of the method of the previous section to define and quantize the higher-spin version of JT gravity. Indeed, one advantage of our method is that there are no further technical difficulties in moving from $\mathrm{SU}(2)$ to an arbitrary gauge group.

The quantization scheme is always the same. We start with an $\mathrm{SL}(N,\mathbb{R})$ BF theory with the given boundary term. We are making the implicit assumption, based on \cite{Li:2013rsa, Datta:2021efl}, that analogous steps to those outlined in Sec. \ref{scales} can be performed. We can make the theory supersymmetric on the hemisphere (topologically equivalent to the disk) as explained in Sec \ref{sec2}. At the end of the day, our action reads again
\begin{equation}
	S_{\textup{tot}} = \frac{\beta}{2\pi} \left[-i\int_{ {\mathrm{HS}^2} } d^2x\sqrt{g}\Tr(\sigma f - \frac{1}{2}\bar\lambda\lambda+ D \eta)+\frac{C}{2}\oint_{\partial\mathrm{HS}^2} \mathrm{d}\varphi \; \Tr(\sigma^2)\right]\;,
\end{equation}
Once the supersymmetric formulation is given, we complexify the fields and choose the contour which makes the supersymmetric path integral convergent. This amounts to choosing the gauge group to be $\mathrm{SU}(N)$, rather than $\mathrm{SL}(N, \mathbb{R})$. We conjecture that this procedure reproduces the higher-spin version of JT gravity.
In the following, we shall recover from a supersymmetric localization perspective all the results in the literature for the partition function on the disk \cite{Gonzalez:2018enk, Datta:2021efl, Kruthoff:2022voq} and explicitly show the relations among the different expressions.

\subsection{The exact partition function}
One main advantage of localization is that it does not depend on the specific gauge group. That choice enters only when one has to make explicit the roots or weights of the gauge group appearing in the classical contribution and in the 1-loop determinant. Therefore, we can start directly from the extension of \eqref{integrale4} to a generic group, with $L$ and $\ell$ replaced by their physical values $\beta$ and $C$ respectively, i.e.
\begin{equation}
	\mathrm{Z}_{\text{JT}} = \frac{1}{N!} \int_{\mathfrak{t}} \mathrm{d}\sigma_{0} \; \exp\!\left(-\frac{\beta}{2 C}\oint \mathrm{d}\varphi \ \Tr(\sigma_0^2)\right) \; \prod_{\alpha>0} [2 \alpha(\sigma_0) \, \sinh(\pi \alpha (\sigma_0))] \;,
\end{equation}
where now the roots $\alpha$ are those of $\mathrm{SU}(N)$. If we use their explicit expression, we find
\begin{equation}
	\mathrm{Z}_{\text{JT}} = \frac{1}{N!} \int \prod_{i}\mathrm{d}\sigma_i \; \delta\bigg(\sum_i\sigma_i\bigg) \, \prod_{i<j}[2(\sigma_i-\sigma_j)\,\sinh(\pi(\sigma_i-\sigma_j))] \; \exp\!\bigg(-\frac{\beta}{2 C}\sum_i\sigma_i^2\bigg)\,.
\end{equation}
where $\sigma_i$, $i=1,\dots,N$ are the eigenvalues of the constant matrix $\sigma_0$, and the delta function set to zero the trace of $\sigma_0$.

We aim to evaluate the integral, which computes the partition function of the $\mathrm{SL}(N, \mathbb{R})$ higher-spin JT gravity on the disk.
Using the integral representation of the Dirac and the Weyl denominator formula\footnote{
	We use the two following variants of the formula
	\begin{align}
		\prod_{1<i<j<N}2\sinh(\pi(\sigma_i-\sigma_j))=\sum_{\eta\in S_N}(-1)^\eta\prod_i e^{2\pi\left(\frac{N+1}{2}-\eta(i)\right)\sigma_i}\,,\\
		\prod_{1<i<j<N}\sigma_j-\sigma_i=\sum_{\lambda\in S_N}(-1)^\lambda\sigma_1^{\lambda(1)-1}\dots\sigma_N^{\lambda(N)-1}=\sum_{\lambda\in S_N}(-1)^\lambda\prod_i\sigma_i^{\lambda(i)-1}\,,
	\end{align}
	where $S_N$ denotes the set of permutations of $N$ elements.
} we arrive at the expression
\begin{align}
	\mathrm{Z}_{\text{JT}}
	= \frac{(-1)^{\frac{N(N-1)}{2}}}{N!} \int \frac{\mathrm{d}k}{2\pi} {}&\sum_{\eta,\lambda\in S_N} (-1)^{\lambda+\eta} \; \prod_i \frac{1}{(2\pi)^{\lambda(i)-1}} \cr
	&\times \frac{\partial}{\partial u^{\lambda(i)-1}_{i}} \int \mathrm{d}\sigma_i \; e^{2\pi\left(\frac{N+1}{2}-\eta(i)+\frac{ik}{2\pi}+u_i\right)\sigma_i-\frac{\beta}{2C}\sigma_i^2} \, \bigg|_{u_i=0} \;,
\end{align}
where we introduced sources $u_i$ to perform the Gaussian integrals.

We can now perform both integrals and obtain
\begin{align}\label{EQ:Z_JT_higher_spin_as_derivative}
	\mathrm{Z}_{\text{JT}}
	= {}&\frac{(-1)^{\frac{N(N-1)}{2}}}{N!\sqrt{N}} \, \bigg(\frac{2\pi C}{\beta}\bigg)^{\!\frac{N-1}{2}} e^{\frac{\pi^2CN(N^2-1)}{6\beta}} \sum_{\lambda,\eta\in S_N} \!(-1)^{\eta+\lambda} \, \prod_i\frac{1}{(2\pi)^{\lambda(i)-1}} \cr
	&\!\!\!\!\! \times \frac{\partial}{\partial u^{\lambda(i)-1}_{i}} \exp\!\bigg(\frac{2\pi^2C}{\beta} \bigg(\!\sum_ju_j^2-\frac{1}{N}\sum_{j,k}u_ju_k+(N+1)\sum_ju_j-2\sum_j\eta(j)u_j\bigg)\!\bigg)\bigg|_{u_i=0} . \cr
\end{align}
We argue that the quadratic part in $u_i$ does not contribute to the final result. To show this, we use the Weyl denominator formula in the opposite direction for the sum over $\eta$ and restore the product of hyperbolic sines:
\begin{multline}
	\sum_{\eta\in S_N}(-1)^\eta \exp\bigg(\frac{2\pi^2C}{\beta} \bigg(\sum_ju_j^2-\frac{1}{N}\sum_{j,k}u_ju_k+(N+1)\sum_ju_j-2\sum_j\eta(j)u_j\bigg)\bigg) \\
	= \exp\bigg(\frac{2\pi^2C}{\beta} \bigg(\sum_ju_j^2-\frac{1}{N}\sum_{j,k}u_ju_k\bigg)\bigg) \; \prod_{i<j}2\sinh\bigg(\frac{2\pi^2 C}{\beta}(u_i-u_j)\bigg) \;.
\end{multline}
To obtain a non-vanishing result, the action of the derivatives on the second factor must act an all the hyperbolic sines. But this can only occur when no derivative acts on the exponential term. As a consequence, we can safely drop the latter from \eqref{EQ:Z_JT_higher_spin_as_derivative}.

The $\beta$-dependence follows from a straightforward scaling argument. If we rename
\begin{equation}
	x_i = \frac{2\pi^2 C}{\beta} \, u_i
\end{equation}
we can easily extract the remaining $\beta$-dependence, that is
\begin{align}
	\mathrm{Z}_{\text{JT}}
	= {}&\frac{(-1)^{\frac{N(N-1)}{2}}}{N!\sqrt{N}} \bigg(\frac{2\pi C}{\beta}\bigg)^{\!\frac{N^2-1}{2}} e^{\frac{\pi^2CN(N^2-1)}{6\beta}} \; \mathcal{K}
\end{align} 
where $\mathcal{K}$ is an overall normalization given by
\begin{align}
	\mathcal{K}
	&= 2^{-\frac{N(N-1)}{2}} \sum_{\eta,\lambda\in S_N} (-1)^{\eta+\lambda} \; \prod_i \frac{\partial}{\partial x^{\lambda(i)-1}_i} \, e^{(N+1)\sum_jx_j-2\sum_j\eta(j)x_j}\,\bigg|_{x_i=0} \cr
	&= 2^{-\frac{N(N-1)}{2}} \sum_{\eta,\lambda\in S_N}(-1)^{\eta+\lambda} \; \prod_i \, [N+1-2\eta(i)]^{\lambda(i)-1} \;.
\end{align}
We get rid of one sum over the permutations by renaming $i\to\eta^{-1}(i)$ and $\pi=\lambda\circ\eta^{-1}$. This gives
\begin{align}
	\mathcal{K}
	&= 2^{-\frac{N(N-1)}{2}} \sum_{\pi\in S_N}(-1)^{\pi}\prod_i  \, [N+1-2\eta(i)]^{\pi(i)-1} \cr
	&= N! \prod_{i<j} (i-j) \cr
	&= (-1)^{\frac{N^2-N}{2}} \, G(N+2) \;,
\end{align}
where $G(x)$ is the Barnes function. In summary, we find
\begin{equation}
	Z_\mathrm{JT} = \frac{G(N+2)}{N!\sqrt{N}} \bigg(\frac{2\pi C}{\beta}\bigg)^{\frac{N^2-1}{2}} \exp\bigg(\pi^2C\frac{N(N^2-1)}{6\beta}\bigg) \;.
\end{equation}
The $\beta$-dependent part agrees nicely with the results obtained by a different localization scheme in \cite{Kruthoff:2022voq}, up to the identification $C = 2\gamma$.  
This computation provides additional and non-trivial evidence that our quantization procedure is not only self-consistent but also suggests an alternative and novel computational method in the context of JT gravity and its generalizations.

\section{Conclusions and Outlooks}\label{concl}
This paper proposes a localization procedure for JT gravity and its higher-spin generalization on the disk topology. We have used a supersymmetric completion of the related gauge theory, involving only auxiliary fields, and complexified the path integration to reduce the computation to a ``standard'' BF theory on the hemisphere.

The correct $s \sinh (2\pi s)$ measure for JT gravity, which in the previous formulations was obtained as the Plancherel measure associated either with the positive semigroup $\mathrm{SL}^{+}(2,\mathbb{R})$ \cite{Blommaert:2018iqz} or with the analytic continuation of the universal cover of $\mathrm{SL}(2,\mathbb{R})$ \cite{Iliesiu:2019xuh}, in our framework is directly provided by gaussian integral over quadratic fluctuations around the dominant saddle point as $t\rightarrow +\infty$. The supersymmetric boundary conditions play a crucial role in annihilating any discrete sum within the moduli space of the localized theory, with the constant configurations of the field $\sigma$ being the only locus to integrate over.

Furthermore, supersymmetry provided us with a crucial boundary potential, quadratic in $\sigma$, which carries the information of the gravitational Gibbons-Hawking term: by carefully establishing the identification between physical and geometrical scales, we have recovered the well-known partition function of JT gravity and confirmed the results \cite{Gonzalez:2018enk,Kruthoff:2022voq} for the higher-spin theory.

The natural follow-up of this work is to generalize the supersymmetric localization of boundary-anchored Wilson lines correlation functions: they admit a simple representation at the level of gauge theory and correspond to correlators of bi-local operators in the boundary Schwarzian quantum mechanics \cite{Blommaert:2018oro,Iliesiu:2019xuh}. The explicit expression for the two-point and the four-point functions has appeared in \cite{Mertens:2017mtv,Blommaert:2018oro,Iliesiu:2019xuh} and was checked to be consistent with direct Schwarzian calculations \cite{Griguolo:2021zsn}. It would be nice to reproduce the result of \cite{Blommaert:2018oro,Iliesiu:2019xuh} from the localization perspective: we expect that representing the Wilson lines would need an extension of the field content of the original BF, probably involving the presence of a chiral multiplet.
If working in the JT gravity case, the procedure could be easily extended to the higher-spin generalization. In this case, exact results are more difficult to extract compared to the purely gravitational case. For instance, if one insists on computing them from the generalized BF approach of \cite{Blommaert:2018oro, Iliesiu:2019xuh}, the complication comes from the lack of explicit expressions for the representation matrices of $\mathrm{SL}(N;\mathbb{R})$.\footnote{See nonetheless \cite{Blommaert:2018oro} for some progress with a focus on the spin-3 case where one can rely on the results of \cite{chervov1999raising}.} Perhaps our localization approach can be used to sidestep such difficulties. Another issue would be to apply our machinery to super-JT gravity, where supersymmetric localization should work along similar lines.

Moreover, JT gravity is known to emerge in the near-horizon limit of four-dimensional extremal black-holes \cite{Iliesiu:2020qvm} and recently supersymmetric localization has been applied to the computation of their entropy \cite{Iliesiu:2022kny}. In light of these advances, it would be interesting to perform the localization of JT gravity in the metric variables in the spirit of analogous higher-dimensional cases \cite{Dabholkar:2011ec}.

As a final comment, although our computation has been performed on a disk topology, JT gravity is known to admit a celebrated non-perturbative completion as a sum over different topologies \cite{Saad:2019lba}. It would be tempting to extend our BF gauge-theoretic approach to higher genus/multi-boundary surfaces. Usually, however, one is faced with the issue that the mapping class group is not taken into account in the BF formulation.\footnote{We thank T. G. Mertens for pointing out this aspect to us.} An easier case where to perform a bulk localization should be given by the singular disk geometry (i.e.\ the ``trumpet''), realized at the gauge-theory level by the insertion of vortex configuration \cite{Hosomichi:2017dbc}.

\acknowledgments{
We thank Marisa Bonini for participating to the early stages of this work, Itamar Yaakov for interesting discussions and useful insights, and Thomas Mertens for reading the manuscript and providing useful suggestions. This work has been supported in part by the Italian Ministero dell’ Universit\`a e Ricerca (MIUR), and by Istituto Nazionale di Fisica Nucleare (INFN) through the ``Gauge and String Theory'' (GAST) research project.}

\appendix
 \addcontentsline{toc}{section}{Appendices}
\begin{flushleft}
\Large\bf Appendices
\end{flushleft}
\section{Conventions}\label{app1}

We recall the conventions for spinors. A spinor $\psi_\alpha$ is a two components column vector. Indices are raised and lowered according to $\psi^\alpha=\epsilon^{\alpha\beta}\psi_\beta$,  $\psi_\alpha=\epsilon_{\alpha\beta}\psi^\beta$, where
\begin{align}
	\epsilon^{\alpha\beta} &= \begin{pmatrix} 0&1\\-1&0\end{pmatrix} \;, &
	\epsilon_{\alpha\beta} &= \begin{pmatrix} 0&-1\\1&0\end{pmatrix} \;.
\end{align}
Standard bilinears are
\begin{align}
	\psi\chi &\equiv \psi^\alpha\chi_\alpha \;, &
	\psi\gamma_\mu\chi &\equiv \psi^\alpha{(\gamma_\mu)_\alpha}^\beta\chi_\beta \;.
\end{align}
Notice that
\begin{align}
	\psi\chi &= (-1)^{h+1}\,\chi\psi \;, &
	\psi\gamma_\mu\chi &= (-1)^h\,\chi\gamma_\mu\psi \;,
\end{align}
where $h=1$ if both $\psi$ and $\chi$ are odd, otherwise $h=0$.

The (flat) gamma matrices satisfy the relations
\begin{align}
	\gamma_a\gamma_b=\delta_{ab}+i\epsilon_{ab}\gamma_3\,,
	\qquad
	\gamma_3\gamma_a=i\epsilon_{ab}\gamma^b
\end{align}
Usefull Fierz identities can be derived just by reducing those for 3d spinors. The basic Fierz identity is 
\begin{align}\label{Fierz}
	\chi_\alpha\psi^\beta=\frac{(-1)^h}{2}\left[\delta_\alpha^\beta(\psi\chi)+(\psi\gamma_a\chi){(\gamma^a)_\alpha}^\beta+(\psi\gamma_3\chi){(\gamma_3)_\alpha}^\beta\right]\,.
\end{align}
For instance, for any spinors $\chi$, $\psi$, and $\lambda$
\begin{align}
	\chi_\alpha(\psi\lambda)+\psi_\alpha(\lambda\chi)+ \lambda_\alpha(\chi\psi)=0\,.
\end{align}

\section{SUSY on the hemisphere}\label{SUSYonHS2}
The metric of the hemisphere reads
\begin{equation}
	\mathrm{d}s^2 = \ell^2\left(\mathrm{d}\theta^2 + \sin^2\!\theta \; \mathrm{d}\varphi^2\right) \;,
\end{equation}
where $\theta\in [0,\frac{\pi}{2}]$ and $\varphi\in [0,2\pi)$, while as vielbein we choose
\begin{align}
	e^{\mathsf{1}} &= \ell \, \mathrm{d}\theta \;, &
	e^{\mathsf{2}} &= \ell \, \sin\theta \, \mathrm{d}\varphi \;.
\end{align}
We also choose $\gamma_{\mathsf{1}}=\sigma_1 $, $\gamma_{\mathsf{2}}=\sigma_2$ and $\gamma_3=\sigma_3$. The spin connection is $\omega_{\mathsf{12}}=-\cos\theta \mathrm{d}\varphi$.

We need to describe only the $\mathcal{N}=(2,2)$ vector multiplet. Its components are the gauge field $A_\mu$, two dimension 1 scalars $\eta $ and $\sigma $, two Dirac fermions $\lambda$, $\bar\lambda$, and the auxiliary field $D$. The corresponding supersymmetry variations are\footnote{To match the notation of \cite{Honda:2013uca}, one needs to identify $\sigma_1 = \eta$ and $\sigma_2 = \sigma$.}
\begin{subequations}
\begin{align}
\delta_{\epsilon,\bar\epsilon} \, A_\mu
	&= -\frac{i}{2}\left(\bar\epsilon\gamma_\mu\lambda+\bar\lambda\gamma_\mu\epsilon\right) \;, \\
\delta_{\epsilon,\bar\epsilon} \, \eta
	&= \frac{1}{2}\left(\bar\epsilon\lambda+\bar\lambda\epsilon\right) \;, \\
\delta_{\epsilon,\bar\epsilon} \, \sigma
	&= -\frac{i}{2}\left(\bar\epsilon\gamma_3\lambda+\bar\lambda\gamma_3\epsilon\right) \;, \\
\delta_{\epsilon,\bar\epsilon} \, \lambda
	&= \left(+i\slashed{\mathrm{D}}\eta + i f \gamma_3 - [\eta,\sigma]\gamma_3 + \frac{i}{\ell}\,\eta\gamma_3 - D + i\epsilon_{\mu\nu}\gamma^\mu\mathrm{D}^\nu\sigma \right)\epsilon \;, \\
\delta_{\epsilon,\bar\epsilon} \, \bar\lambda
	&= \left(-i\slashed{\mathrm{D}}\eta + i f \gamma_3 + [\eta,\sigma]\gamma_3 + \frac{i}{\ell}\,\eta\gamma_3 + D + i\epsilon_{\mu\nu}\gamma^\mu\mathrm{D}^\nu\sigma \right)\bar\epsilon \;, \\
\delta_{\epsilon,\bar\epsilon} \, D
	&= -\frac{i}{2}\,\bar\epsilon \slashed{\mathrm{D}}\lambda - \frac{i}{2}\,[\eta,\bar\epsilon\lambda] - \frac{1}{2}\,[\sigma,\bar\epsilon\gamma_3\lambda] + \frac{i}{2}\,\epsilon\slashed{\mathrm{D}}\bar\lambda + \frac{i}{2}\,[\eta,\bar\lambda\epsilon] + \frac{1}{2}\,[\sigma,\bar\lambda\gamma_3\epsilon] \;.
\end{align}
\end{subequations}
$\epsilon$ and $\bar\epsilon$ are bosonic spinors satisfying the conformal Killing spinor equation 
\begin{equation}
	\nabla_\mu \epsilon=\gamma_\mu\tilde\epsilon\,,
\end{equation}
for some spinor $\tilde\epsilon$. It is solved by 
\begin{align}
	\epsilon &= e^{-s\frac{i}{2}\theta\gamma^{\mathsf{2}}}\begin{pmatrix} e^{\frac{i}{2}\varphi}\\0\end{pmatrix} &
	&\text{and}&
	\epsilon &= e^{-s\frac{i}{2}\theta\gamma^{\mathsf{2}}}\begin{pmatrix} 0\\e^{-\frac{i}{2}\varphi}\end{pmatrix} \;,
\end{align}
where $s=\pm1$. The same solutions hold for $\bar\epsilon$.

Together, $\epsilon$ and $\bar\epsilon$ generate the entire 2d superconformal algebra. To apply localization on $S^2$, we can restrict to the $\mathfrak{su}(2|1)$ Poincar\'e subalgebra, generated by the only four Killing spinors
\begin{align}\label{possible}
	\epsilon &= e^{-\frac{i}{2}\theta\gamma^{\mathsf{2}}} \begin{pmatrix} e^{\frac{i}{2}\varphi}\\0\end{pmatrix} \;, &
	\epsilon &= e^{+\frac{i}{2}\theta\gamma^{\mathsf{2}}} \begin{pmatrix} 0\\e^{-\frac{i}{2}\varphi}\end{pmatrix} \;, \cr
	\bar\epsilon &= e^{+\frac{i}{2}\theta\gamma^{\mathsf{2}}} \begin{pmatrix} e^{\frac{i}{2}\varphi}\\0\end{pmatrix} \;, &
	\bar\epsilon &= e^{-\frac{i}{2}\theta\gamma^{\mathsf{2}}} \begin{pmatrix} 0\\e^{-\frac{i}{2}\varphi}\end{pmatrix} \;.
\end{align}

They satisfy the equations
\begin{align}
	\nabla_\mu\epsilon &= \frac{1}{2\ell}\gamma_\mu\gamma_3\epsilon \;, &
	\nabla_\mu\bar\epsilon &= -\frac{1}{2\ell}\gamma_\mu\gamma_3\bar\epsilon \;.
\end{align}
We have four solutions. In restricting to the hemisphere, the $\mathrm{SU}(2)$ isometry group breaks down to the $\mathrm{U}(1)$ group of azimuthal rotations. Therefore, we expect that supersymmetry will close on a $\mathfrak{su}(1|1)$ algebra, with spacetime symmetry reduced to rotations along $\varphi$. We choose the first couple in \eqref{possible} 
\begin{align}\label{preservedsusy}
\epsilon &= e^{\frac{i \varphi }{2}} \begin{pmatrix}\cos \frac{\theta }{2}\\\sin \frac{\theta }{2}\end{pmatrix} \;,
&
\bar\epsilon &= e^{-\frac{i \varphi }{2} }\begin{pmatrix}  \sin \frac{\theta }{2}\\ \cos \frac{\theta }{2} \end{pmatrix} \;.
\end{align}

Some useful Killing spinor bilinears are
\begin{align}
	\bar\epsilon\gamma^3\epsilon &= 1 \;, &
	\bar\epsilon\epsilon &= \cos\theta \;, &
	\bar\epsilon\gamma^\mu\epsilon &= \left(0,-\frac i\ell\right) \;.
\end{align}

\bibliography{biblio}

\providecommand{\href}[2]{#2}\begingroup\raggedright\begin{thebibliography}{10}

\bibitem{Maldacena:1997re}
J.~M. Maldacena, \emph{{The Large N limit of superconformal field theories and supergravity}}, \href{https://doi.org/10.4310/ATMP.1998.v2.n2.a1}{\emph{Adv. Theor. Math. Phys.} {\bfseries 2} (1998) 231} [\href{https://arxiv.org/abs/hep-th/9711200}{{\ttfamily hep-th/9711200}}].

\bibitem{Witten:1998qj}
E.~Witten, \emph{{Anti-de Sitter space and holography}}, \href{https://doi.org/10.4310/ATMP.1998.v2.n2.a2}{\emph{Adv. Theor. Math. Phys.} {\bfseries 2} (1998) 253} [\href{https://arxiv.org/abs/hep-th/9802150}{{\ttfamily hep-th/9802150}}].

\bibitem{Gubser:1998bc}
S.~S. Gubser, I.~R. Klebanov and A.~M. Polyakov, \emph{{Gauge theory correlators from noncritical string theory}}, \href{https://doi.org/10.1016/S0370-2693(98)00377-3}{\emph{Phys. Lett. B} {\bfseries 428} (1998) 105} [\href{https://arxiv.org/abs/hep-th/9802109}{{\ttfamily hep-th/9802109}}].

\bibitem{Pestun:2016zxk}
V.~Pestun et~al., \emph{{Localization techniques in quantum field theories}}, \href{https://doi.org/10.1088/1751-8121/aa63c1}{\emph{J. Phys. A} {\bfseries 50} (2017) 440301} [\href{https://arxiv.org/abs/1608.02952}{{\ttfamily 1608.02952}}].

\bibitem{Benini:2015noa}
F.~Benini and A.~Zaffaroni, \emph{{A topologically twisted index for three-dimensional supersymmetric theories}}, \href{https://doi.org/10.1007/JHEP07(2015)127}{\emph{JHEP} {\bfseries 07} (2015) 127} [\href{https://arxiv.org/abs/1504.03698}{{\ttfamily 1504.03698}}].

\bibitem{Benini:2015eyy}
F.~Benini, K.~Hristov and A.~Zaffaroni, \emph{{Black hole microstates in AdS$_{4}$ from supersymmetric localization}}, \href{https://doi.org/10.1007/JHEP05(2016)054}{\emph{JHEP} {\bfseries 05} (2016) 054} [\href{https://arxiv.org/abs/1511.04085}{{\ttfamily 1511.04085}}].

\bibitem{Azzurli:2017kxo}
F.~Azzurli, N.~Bobev, P.~M. Crichigno, V.~S. Min and A.~Zaffaroni, \emph{{A universal counting of black hole microstates in AdS$_{4}$}}, \href{https://doi.org/10.1007/JHEP02(2018)054}{\emph{JHEP} {\bfseries 02} (2018) 054} [\href{https://arxiv.org/abs/1707.04257}{{\ttfamily 1707.04257}}].

\bibitem{Cabo-Bizet:2018ehj}
A.~Cabo-Bizet, D.~Cassani, D.~Martelli and S.~Murthy, \emph{{Microscopic origin of the Bekenstein-Hawking entropy of supersymmetric AdS$_{5}$ black holes}}, \href{https://doi.org/10.1007/JHEP10(2019)062}{\emph{JHEP} {\bfseries 10} (2019) 062} [\href{https://arxiv.org/abs/1810.11442}{{\ttfamily 1810.11442}}].

\bibitem{Choi:2018hmj}
S.~Choi, J.~Kim, S.~Kim and J.~Nahmgoong, \emph{{Large AdS black holes from QFT}},  \href{https://arxiv.org/abs/1810.12067}{{\ttfamily 1810.12067}}.

\bibitem{Benini:2018ywd}
F.~Benini and E.~Milan, \emph{{Black Holes in 4D $\mathcal{N}$=4 Super-Yang-Mills Field Theory}}, \href{https://doi.org/10.1103/PhysRevX.10.021037}{\emph{Phys. Rev. X} {\bfseries 10} (2020) 021037} [\href{https://arxiv.org/abs/1812.09613}{{\ttfamily 1812.09613}}].

\bibitem{Dabholkar:2014wpa}
A.~Dabholkar, N.~Drukker and J.~Gomes, \emph{{Localization in supergravity and quantum $AdS_4/CFT_3$ holography}}, \href{https://doi.org/10.1007/JHEP10(2014)090}{\emph{JHEP} {\bfseries 10} (2014) 090} [\href{https://arxiv.org/abs/1406.0505}{{\ttfamily 1406.0505}}].

\bibitem{Murthy:2015yfa}
S.~Murthy and V.~Reys, \emph{{Functional determinants, index theorems, and exact quantum black hole entropy}}, \href{https://doi.org/10.1007/JHEP12(2015)028}{\emph{JHEP} {\bfseries 12} (2015) 028} [\href{https://arxiv.org/abs/1504.01400}{{\ttfamily 1504.01400}}].

\bibitem{Teitelboim:1983ux}
C.~Teitelboim, \emph{{Gravitation and Hamiltonian Structure in Two Space-Time Dimensions}}, \href{https://doi.org/10.1016/0370-2693(83)90012-6}{\emph{Phys. Lett. B} {\bfseries 126} (1983) 41}.

\bibitem{Jackiw:1984je}
R.~Jackiw, \emph{{Lower Dimensional Gravity}}, \href{https://doi.org/10.1016/0550-3213(85)90448-1}{\emph{Nucl. Phys. B} {\bfseries 252} (1985) 343}.

\bibitem{Mertens:2022irh}
T.~G. Mertens and G.~J. Turiaci, \emph{{Solvable Models of Quantum Black Holes: A Review on Jackiw-Teitelboim Gravity}},  \href{https://arxiv.org/abs/2210.10846}{{\ttfamily 2210.10846}}.

\bibitem{Maldacena:2016upp}
J.~Maldacena, D.~Stanford and Z.~Yang, \emph{{Conformal symmetry and its breaking in two dimensional Nearly Anti-de-Sitter space}}, \href{https://doi.org/10.1093/ptep/ptw124}{\emph{PTEP} {\bfseries 2016} (2016) 12C104} [\href{https://arxiv.org/abs/1606.01857}{{\ttfamily 1606.01857}}].

\bibitem{Stanford:2017thb}
D.~Stanford and E.~Witten, \emph{{Fermionic Localization of the Schwarzian Theory}}, \href{https://doi.org/10.1007/JHEP10(2017)008}{\emph{JHEP} {\bfseries 10} (2017) 008} [\href{https://arxiv.org/abs/1703.04612}{{\ttfamily 1703.04612}}].

\bibitem{Kitaev:2017awl}
A.~Kitaev and S.~J. Suh, \emph{{The soft mode in the Sachdev-Ye-Kitaev model and its gravity dual}}, \href{https://doi.org/10.1007/JHEP05(2018)183}{\emph{JHEP} {\bfseries 05} (2018) 183} [\href{https://arxiv.org/abs/1711.08467}{{\ttfamily 1711.08467}}].

\bibitem{Saad:2018bqo}
P.~Saad, S.~H. Shenker and D.~Stanford, \emph{{A semiclassical ramp in SYK and in gravity}},  \href{https://arxiv.org/abs/1806.06840}{{\ttfamily 1806.06840}}.

\bibitem{Stanford:2019vob}
D.~Stanford and E.~Witten, \emph{{JT gravity and the ensembles of random matrix theory}}, \href{https://doi.org/10.4310/ATMP.2020.v24.n6.a4}{\emph{Adv. Theor. Math. Phys.} {\bfseries 24} (2020) 1475} [\href{https://arxiv.org/abs/1907.03363}{{\ttfamily 1907.03363}}].

\bibitem{Penington:2019kki}
G.~Penington, S.~H. Shenker, D.~Stanford and Z.~Yang, \emph{{Replica wormholes and the black hole interior}}, \href{https://doi.org/10.1007/JHEP03(2022)205}{\emph{JHEP} {\bfseries 03} (2022) 205} [\href{https://arxiv.org/abs/1911.11977}{{\ttfamily 1911.11977}}].

\bibitem{Almheiri:2019qdq}
A.~Almheiri, T.~Hartman, J.~Maldacena, E.~Shaghoulian and A.~Tajdini, \emph{{Replica Wormholes and the Entropy of Hawking Radiation}}, \href{https://doi.org/10.1007/JHEP05(2020)013}{\emph{JHEP} {\bfseries 05} (2020) 013} [\href{https://arxiv.org/abs/1911.12333}{{\ttfamily 1911.12333}}].

\bibitem{Saad:2019lba}
P.~Saad, S.~H. Shenker and D.~Stanford, \emph{{JT gravity as a matrix integral}},  \href{https://arxiv.org/abs/1903.11115}{{\ttfamily 1903.11115}}.

\bibitem{Benini:2012ui}
F.~Benini and S.~Cremonesi, \emph{{Partition Functions of ${\mathcal{N}=(2,2)}$ Gauge Theories on S$^{2}$ and Vortices}}, \href{https://doi.org/10.1007/s00220-014-2112-z}{\emph{Commun. Math. Phys.} {\bfseries 334} (2015) 1483} [\href{https://arxiv.org/abs/1206.2356}{{\ttfamily 1206.2356}}].

\bibitem{Iizuka:2015jma}
N.~Iizuka, A.~Tanaka and S.~Terashima, \emph{{Exact Path Integral for 3D Quantum Gravity}}, \href{https://doi.org/10.1103/PhysRevLett.115.161304}{\emph{Phys. Rev. Lett.} {\bfseries 115} (2015) 161304} [\href{https://arxiv.org/abs/1504.05991}{{\ttfamily 1504.05991}}].

\bibitem{Castro:2023bvo}
A.~Castro, I.~Coman, J.~R. Fliss and C.~Zukowski, \emph{{Coupling Fields to 3D Quantum Gravity via Chern-Simons Theory}},  \href{https://arxiv.org/abs/2304.02668}{{\ttfamily 2304.02668}}.

\bibitem{Blommaert:2018oro}
A.~Blommaert, T.~G. Mertens and H.~Verschelde, \emph{{The Schwarzian Theory - A Wilson Line Perspective}}, \href{https://doi.org/10.1007/JHEP12(2018)022}{\emph{JHEP} {\bfseries 12} (2018) 022} [\href{https://arxiv.org/abs/1806.07765}{{\ttfamily 1806.07765}}].

\bibitem{Mertens:2018fds}
T.~G. Mertens, \emph{{The Schwarzian theory \textemdash{} origins}}, \href{https://doi.org/10.1007/JHEP05(2018)036}{\emph{JHEP} {\bfseries 05} (2018) 036} [\href{https://arxiv.org/abs/1801.09605}{{\ttfamily 1801.09605}}].

\bibitem{Iliesiu:2019xuh}
L.~V. Iliesiu, S.~S. Pufu, H.~Verlinde and Y.~Wang, \emph{{An exact quantization of Jackiw-Teitelboim gravity}}, \href{https://doi.org/10.1007/JHEP11(2019)091}{\emph{JHEP} {\bfseries 11} (2019) 091} [\href{https://arxiv.org/abs/1905.02726}{{\ttfamily 1905.02726}}].

\bibitem{Alkalaev:2014qpa}
K.~B. Alkalaev, \emph{{Global and local properties of AdS$_{2}$ higher spin gravity}}, \href{https://doi.org/10.1007/JHEP10(2014)122}{\emph{JHEP} {\bfseries 10} (2014) 122} [\href{https://arxiv.org/abs/1404.5330}{{\ttfamily 1404.5330}}].

\bibitem{Gonzalez:2018enk}
H.~A. Gonz\'alez, D.~Grumiller and J.~Salzer, \emph{{Towards a bulk description of higher spin SYK}}, \href{https://doi.org/10.1007/JHEP05(2018)083}{\emph{JHEP} {\bfseries 05} (2018) 083} [\href{https://arxiv.org/abs/1802.01562}{{\ttfamily 1802.01562}}].

\bibitem{Datta:2021efl}
S.~Datta, \emph{{The Schwarzian sector of higher spin CFTs}}, \href{https://doi.org/10.1007/JHEP04(2021)171}{\emph{JHEP} {\bfseries 04} (2021) 171} [\href{https://arxiv.org/abs/2101.04980}{{\ttfamily 2101.04980}}].

\bibitem{Kruthoff:2022voq}
J.~Kruthoff, \emph{{Higher spin JT gravity and a matrix model dual}}, \href{https://doi.org/10.1007/JHEP09(2022)017}{\emph{JHEP} {\bfseries 09} (2022) 017} [\href{https://arxiv.org/abs/2204.09685}{{\ttfamily 2204.09685}}].

\bibitem{Mertens:2017mtv}
T.~G. Mertens, G.~J. Turiaci and H.~L. Verlinde, \emph{{Solving the Schwarzian via the Conformal Bootstrap}}, \href{https://doi.org/10.1007/JHEP08(2017)136}{\emph{JHEP} {\bfseries 08} (2017) 136} [\href{https://arxiv.org/abs/1705.08408}{{\ttfamily 1705.08408}}].

\bibitem{Fukuyama:1985gg}
T.~Fukuyama and K.~Kamimura, \emph{{Gauge Theory of Two-dimensional Gravity}}, \href{https://doi.org/10.1016/0370-2693(85)91322-X}{\emph{Phys. Lett. B} {\bfseries 160} (1985) 259}.

\bibitem{Isler:1989hq}
K.~Isler and C.~A. Trugenberger, \emph{{A Gauge Theory of Two-dimensional Quantum Gravity}}, \href{https://doi.org/10.1103/PhysRevLett.63.834}{\emph{Phys. Rev. Lett.} {\bfseries 63} (1989) 834}.

\bibitem{Chamseddine:1989yz}
A.~H. Chamseddine and D.~Wyler, \emph{{Gauge Theory of Topological Gravity in (1+1)-Dimensions}}, \href{https://doi.org/10.1016/0370-2693(89)90528-5}{\emph{Phys. Lett. B} {\bfseries 228} (1989) 75}.

\bibitem{Sugishita:2013jca}
S.~Sugishita and S.~Terashima, \emph{{Exact Results in Supersymmetric Field Theories on Manifolds with Boundaries}}, \href{https://doi.org/10.1007/JHEP11(2013)021}{\emph{JHEP} {\bfseries 11} (2013) 021} [\href{https://arxiv.org/abs/1308.1973}{{\ttfamily 1308.1973}}].

\bibitem{Honda:2013uca}
D.~Honda and T.~Okuda, \emph{{Exact results for boundaries and domain walls in 2d supersymmetric theories}}, \href{https://doi.org/10.1007/JHEP09(2015)140}{\emph{JHEP} {\bfseries 09} (2015) 140} [\href{https://arxiv.org/abs/1308.2217}{{\ttfamily 1308.2217}}].

\bibitem{Hori:2013ika}
K.~Hori and M.~Romo, \emph{{Exact Results In Two-Dimensional (2,2) Supersymmetric Gauge Theories With Boundary}},  \href{https://arxiv.org/abs/1308.2438}{{\ttfamily 1308.2438}}.

\bibitem{Banados:1992wn}
M.~Banados, C.~Teitelboim and J.~Zanelli, \emph{{The Black hole in three-dimensional space-time}}, \href{https://doi.org/10.1103/PhysRevLett.69.1849}{\emph{Phys. Rev. Lett.} {\bfseries 69} (1992) 1849} [\href{https://arxiv.org/abs/hep-th/9204099}{{\ttfamily hep-th/9204099}}].

\bibitem{Banados:1992gq}
M.~Banados, M.~Henneaux, C.~Teitelboim and J.~Zanelli, \emph{{Geometry of the (2+1) black hole}}, \href{https://doi.org/10.1103/PhysRevD.48.1506}{\emph{Phys. Rev. D} {\bfseries 48} (1993) 1506} [\href{https://arxiv.org/abs/gr-qc/9302012}{{\ttfamily gr-qc/9302012}}].

\bibitem{Mertens:2022ujr}
T.~G. Mertens, J.~Sim\'on and G.~Wong, \emph{{A proposal for 3d quantum gravity and its bulk factorization}},  \href{https://arxiv.org/abs/2210.14196}{{\ttfamily 2210.14196}}.

\bibitem{Jensen:2016pah}
K.~Jensen, \emph{{Chaos in AdS$_2$ Holography}}, \href{https://doi.org/10.1103/PhysRevLett.117.111601}{\emph{Phys. Rev. Lett.} {\bfseries 117} (2016) 111601} [\href{https://arxiv.org/abs/1605.06098}{{\ttfamily 1605.06098}}].

\bibitem{Engelsoy:2016xyb}
J.~Engels\"oy, T.~G. Mertens and H.~Verlinde, \emph{{An investigation of AdS$_{2}$ backreaction and holography}}, \href{https://doi.org/10.1007/JHEP07(2016)139}{\emph{JHEP} {\bfseries 07} (2016) 139} [\href{https://arxiv.org/abs/1606.03438}{{\ttfamily 1606.03438}}].

\bibitem{Cotler:2018zff}
J.~Cotler and K.~Jensen, \emph{{A theory of reparameterizations for AdS$_3$ gravity}}, \href{https://doi.org/10.1007/JHEP02(2019)079}{\emph{JHEP} {\bfseries 02} (2019) 079} [\href{https://arxiv.org/abs/1808.03263}{{\ttfamily 1808.03263}}].

\bibitem{Witten:1988hc}
E.~Witten, \emph{{(2+1)-Dimensional Gravity as an Exactly Soluble System}}, \href{https://doi.org/10.1016/0550-3213(88)90143-5}{\emph{Nucl. Phys. B} {\bfseries 311} (1988) 46}.

\bibitem{Achucarro:1986uwr}
A.~Achucarro and P.~K. Townsend, \emph{{A Chern-Simons Action for Three-Dimensional anti-De Sitter Supergravity Theories}}, \href{https://doi.org/10.1016/0370-2693(86)90140-1}{\emph{Phys. Lett. B} {\bfseries 180} (1986) 89}.

\bibitem{Pestun:2007rz}
V.~Pestun, \emph{{Localization of gauge theory on a four-sphere and supersymmetric Wilson loops}}, \href{https://doi.org/10.1007/s00220-012-1485-0}{\emph{Commun. Math. Phys.} {\bfseries 313} (2012) 71} [\href{https://arxiv.org/abs/0712.2824}{{\ttfamily 0712.2824}}].

\bibitem{Kapustin:2009kz}
A.~Kapustin, B.~Willett and I.~Yaakov, \emph{{Exact Results for Wilson Loops in Superconformal Chern-Simons Theories with Matter}}, \href{https://doi.org/10.1007/JHEP03(2010)089}{\emph{JHEP} {\bfseries 03} (2010) 089} [\href{https://arxiv.org/abs/0909.4559}{{\ttfamily 0909.4559}}].

\bibitem{Fradkin:1987ks}
E.~S. Fradkin and M.~A. Vasiliev, \emph{{On the Gravitational Interaction of Massless Higher Spin Fields}}, \href{https://doi.org/10.1016/0370-2693(87)91275-5}{\emph{Phys. Lett. B} {\bfseries 189} (1987) 89}.

\bibitem{Vasiliev:1990en}
M.~A. Vasiliev, \emph{{Consistent equation for interacting gauge fields of all spins in (3+1)-dimensions}}, \href{https://doi.org/10.1016/0370-2693(90)91400-6}{\emph{Phys. Lett. B} {\bfseries 243} (1990) 378}.

\bibitem{Vasiliev:1992av}
M.~A. Vasiliev, \emph{{More on equations of motion for interacting massless fields of all spins in (3+1)-dimensions}}, \href{https://doi.org/10.1016/0370-2693(92)91457-K}{\emph{Phys. Lett. B} {\bfseries 285} (1992) 225}.

\bibitem{Vasiliev:1995dn}
M.~A. Vasiliev, \emph{{Higher spin gauge theories in four-dimensions, three-dimensions, and two-dimensions}}, \href{https://doi.org/10.1142/S0218271896000473}{\emph{Int. J. Mod. Phys. D} {\bfseries 5} (1996) 763} [\href{https://arxiv.org/abs/hep-th/9611024}{{\ttfamily hep-th/9611024}}].

\bibitem{Vasiliev:1999ba}
M.~A. Vasiliev, \emph{{Higher spin gauge theories: Star product and AdS space}},  \href{https://arxiv.org/abs/hep-th/9910096}{{\ttfamily hep-th/9910096}}.

\bibitem{Vasiliev:2003ev}
M.~A. Vasiliev, \emph{{Nonlinear equations for symmetric massless higher spin fields in (A)dS(d)}}, \href{https://doi.org/10.1016/S0370-2693(03)00872-4}{\emph{Phys. Lett. B} {\bfseries 567} (2003) 139} [\href{https://arxiv.org/abs/hep-th/0304049}{{\ttfamily hep-th/0304049}}].

\bibitem{Klebanov:2002ja}
I.~R. Klebanov and A.~M. Polyakov, \emph{{AdS dual of the critical O(N) vector model}}, \href{https://doi.org/10.1016/S0370-2693(02)02980-5}{\emph{Phys. Lett. B} {\bfseries 550} (2002) 213} [\href{https://arxiv.org/abs/hep-th/0210114}{{\ttfamily hep-th/0210114}}].

\bibitem{Gaberdiel:2010pz}
M.~R. Gaberdiel and R.~Gopakumar, \emph{{An AdS$_{3}$ Dual for Minimal Model CFTs}}, \href{https://doi.org/10.1103/PhysRevD.83.066007}{\emph{Phys. Rev. D} {\bfseries 83} (2011) 066007} [\href{https://arxiv.org/abs/1011.2986}{{\ttfamily 1011.2986}}].

\bibitem{Blencowe:1988gj}
M.~P. Blencowe, \emph{{A Consistent Interacting Massless Higher Spin Field Theory in $D$ = (2+1)}}, \href{https://doi.org/10.1088/0264-9381/6/4/005}{\emph{Class. Quant. Grav.} {\bfseries 6} (1989) 443}.

\bibitem{Aragone:1983sz}
C.~Aragone and S.~Deser, \emph{{Hypersymmetry in $D=3$ of Coupled Gravity Massless Spin 5/2 System}}, \href{https://doi.org/10.1088/0264-9381/1/2/001}{\emph{Class. Quant. Grav.} {\bfseries 1} (1984) L9}.

\bibitem{Campoleoni:2010zq}
A.~Campoleoni, S.~Fredenhagen, S.~Pfenninger and S.~Theisen, \emph{{Asymptotic symmetries of three-dimensional gravity coupled to higher-spin fields}}, \href{https://doi.org/10.1007/JHEP11(2010)007}{\emph{JHEP} {\bfseries 11} (2010) 007} [\href{https://arxiv.org/abs/1008.4744}{{\ttfamily 1008.4744}}].

\bibitem{Campoleoni:2011hg}
A.~Campoleoni, S.~Fredenhagen and S.~Pfenninger, \emph{{Asymptotic W-symmetries in three-dimensional higher-spin gauge theories}}, \href{https://doi.org/10.1007/JHEP09(2011)113}{\emph{JHEP} {\bfseries 09} (2011) 113} [\href{https://arxiv.org/abs/1107.0290}{{\ttfamily 1107.0290}}].

\bibitem{Alkalaev:2013fsa}
K.~B. Alkalaev, \emph{{On higher spin extension of the Jackiw-Teitelboim gravity model}}, \href{https://doi.org/10.1088/1751-8113/47/36/365401}{\emph{J. Phys. A} {\bfseries 47} (2014) 365401} [\href{https://arxiv.org/abs/1311.5119}{{\ttfamily 1311.5119}}].

\bibitem{Grumiller:2013swa}
D.~Grumiller, M.~Leston and D.~Vassilevich, \emph{{Anti-de Sitter holography for gravity and higher spin theories in two dimensions}}, \href{https://doi.org/10.1103/PhysRevD.89.044001}{\emph{Phys. Rev. D} {\bfseries 89} (2014) 044001} [\href{https://arxiv.org/abs/1311.7413}{{\ttfamily 1311.7413}}].

\bibitem{Narayan:2019ove}
P.~Narayan and J.~Yoon, \emph{{Chaos in Three-dimensional Higher Spin Gravity}}, \href{https://doi.org/10.1007/JHEP07(2019)046}{\emph{JHEP} {\bfseries 07} (2019) 046} [\href{https://arxiv.org/abs/1903.08761}{{\ttfamily 1903.08761}}].

\bibitem{Bergshoeff:1989ns}
E.~Bergshoeff, M.~P. Blencowe and K.~S. Stelle, \emph{{Area Preserving Diffeomorphisms and Higher Spin Algebra}}, \href{https://doi.org/10.1007/BF02108779}{\emph{Commun. Math. Phys.} {\bfseries 128} (1990) 213}.

\bibitem{Li:2013rsa}
W.~Li, F.-L. Lin and C.-W. Wang, \emph{{Modular Properties of 3D Higher Spin Theory}}, \href{https://doi.org/10.1007/JHEP12(2013)094}{\emph{JHEP} {\bfseries 12} (2013) 094} [\href{https://arxiv.org/abs/1308.2959}{{\ttfamily 1308.2959}}].

\bibitem{Blommaert:2018iqz}
A.~Blommaert, T.~G. Mertens and H.~Verschelde, \emph{{Fine Structure of Jackiw-Teitelboim Quantum Gravity}}, \href{https://doi.org/10.1007/JHEP09(2019)066}{\emph{JHEP} {\bfseries 09} (2019) 066} [\href{https://arxiv.org/abs/1812.00918}{{\ttfamily 1812.00918}}].

\bibitem{Griguolo:2021zsn}
L.~Griguolo, J.~Papalini and D.~Seminara, \emph{{On the perturbative expansion of exact bi-local correlators in JT gravity}}, \href{https://doi.org/10.1007/JHEP05(2021)140}{\emph{JHEP} {\bfseries 05} (2021) 140} [\href{https://arxiv.org/abs/2101.06252}{{\ttfamily 2101.06252}}].

\bibitem{chervov1999raising}
A.~Chervov, \emph{Raising operators for the whittaker wave functions of the toda chain and intertwining operators},  1999.

\bibitem{Iliesiu:2020qvm}
L.~V. Iliesiu and G.~J. Turiaci, \emph{{The statistical mechanics of near-extremal black holes}}, \href{https://doi.org/10.1007/JHEP05(2021)145}{\emph{JHEP} {\bfseries 05} (2021) 145} [\href{https://arxiv.org/abs/2003.02860}{{\ttfamily 2003.02860}}].

\bibitem{Iliesiu:2022kny}
L.~V. Iliesiu, S.~Murthy and G.~J. Turiaci, \emph{{Black hole microstate counting from the gravitational path integral}},  \href{https://arxiv.org/abs/2209.13602}{{\ttfamily 2209.13602}}.

\bibitem{Dabholkar:2011ec}
A.~Dabholkar, J.~Gomes and S.~Murthy, \emph{{Localization \& Exact Holography}}, \href{https://doi.org/10.1007/JHEP04(2013)062}{\emph{JHEP} {\bfseries 04} (2013) 062} [\href{https://arxiv.org/abs/1111.1161}{{\ttfamily 1111.1161}}].

\bibitem{Hosomichi:2017dbc}
K.~Hosomichi, S.~Lee and T.~Okuda, \emph{{Supersymmetric vortex defects in two dimensions}}, \href{https://doi.org/10.1007/JHEP01(2018)033}{\emph{JHEP} {\bfseries 01} (2018) 033} [\href{https://arxiv.org/abs/1705.10623}{{\ttfamily 1705.10623}}].

\end{thebibliography}\endgroup

\end{document}